\documentclass[aps,prb,reprint,superscriptaddress,showpacs]{revtex4-1}
\usepackage[german,english]{babel}
\usepackage{amsmath,amssymb}
\usepackage{amsbsy}
\usepackage{graphicx}

\let\vec=\mathbf
\newcommand{\mean}[1]{\left\langle #1 \right\rangle}
\newcommand{\abs}[1]{\left| #1 \right|}
\newcommand{\calG}{{\cal G}}
\newcommand{\cimp}{c_{\mathrm{imp}}}
\newcommand{\Vimp}{\check{V}_{\mathrm{imp}}}

\newcommand{\Gnaught}{\check{G}_{E}^{(0)}}

\newcommand{\tdep}{\tau_{\mathrm{dep}}}

\def\Xint#1{\mathchoice
   {\XXint\displaystyle\textstyle{#1}}%
   {\XXint\textstyle\scriptstyle{#1}}%
   {\XXint\scriptstyle\scriptscriptstyle{#1}}%
   {\XXint\scriptscriptstyle\scriptscriptstyle{#1}}%
   \!\int}
\def\XXint#1#2#3{{\setbox0=\hbox{$#1{#2#3}{\int}$}
     \vcenter{\hbox{$#2#3$}}\kern-.5\wd0}}

\def\dashint{\Xint-}

\begin{document}

\title{Density of states in gapped superconductors with pairing-potential impurities}
\author{Anton~Bespalov}
\affiliation{Univ.~Grenoble Alpes, INAC-SPSMS, F-38000 Grenoble, France;\\
CEA, INAC-SPSMS, F-38000 Grenoble, France.}
\affiliation{Institute for Physics of Microstructures, Russian Academy of Sciences, 603950 Nizhny Novgorod, GSP-105, Russia}
\author{Manuel~Houzet}
\author{Julia~S.~Meyer}
\affiliation{Univ.~Grenoble Alpes, INAC-SPSMS, F-38000 Grenoble, France;\\
CEA, INAC-SPSMS, F-38000 Grenoble, France.}
\author{Yuli~V.~Nazarov}
\affiliation{Kavli Institute of NanoScience, Delft University of Technology, Lorentzweg 1, NL-2628 CJ, Delft, The Netherlands.}

\begin{abstract}
We study the density of states in disordered s-wave superconductors with a small gap anisotropy. Disorder comes in the form of common nonmagnetic scatterers and pairing-potential impurities, which interact with electrons via an electric potential and a local distortion of the superconducting gap. A set of equations for the quasiclassical Green functions is derived and solved. Within one spin sector, pairing-potential impurities and weak spin-polarized magnetic impurities have essentially the same effect on the density of states. We show that if the gap is isotropic, an isolated impurity with suppressed pairing supports an infinite number of Andreev states. With growing impurity concentration, the energy-dependent density of states evolves from a sharp gap edge with an impurity band below it to a smeared BCS singularity in the so-called universal limit. If a gap anisotropy is present, the density of states becomes sensitive to ordinary potential disorder, and the existence of of Andreev states localized at pairing-potential impurities requires special conditions. An unusual feature related to the anisotropy is a nonmonotonic dependence of the gap edge smearing on impurity concentration.
\end{abstract}
\pacs{74.62.En}

\maketitle

\section{Introduction}
\label{sec:intro}
Thermodynamic and transport properties of disordered superconductors crucially depend on the symmetry of superconducting pairing as well as on the nature of the impurities that scatter the electron waves. It is widely known that conventional scatterers described by coordinate-dependent potentials hardly affect the density of states or the order parameter in superconductors with conventional spin-singlet $s$-wave pairing 
\cite{Anderson1959JPCS,Tsuneto62PTP,Hohenberg64JETP_eng,Clem66PR}. Their main effect is the suppression of the anisotropic part of the order parameter, which is small as far as the anisotropic part is small. By contrast, for unconventional superconducting pairing that is essentially anisotropic, the effect of potential disorder on the density of states is drastic. For instance, even a single potential scatterer in a $d$-wave superconductor brings about a quasibound state localized at the defect\cite{Balatsky+95PRB}, while a large concentration of defects leads to the complete suppression of superconductivity.
 
The situation is different for magnetic impurities \cite{AG61_eng}. In an $s$-wave 
superconductor, a single magnetic impurity induces a localized state, known as a Yu\cite{Yu1965}, Shiba\cite{Shiba68PTP}, or Rusinov\cite{Rusinov1969JETP} state, with an energy below the gap edge. If the exchange field of the magnetic impurity is weak, the impurity state is formed close to the gap edge.  At finite impurity concentration, $\cimp$, the Shiba states hybridize and form an impurity  band that becomes wider with increasing $\cimp$. 
Simultaneously, the disorder smears the BCS singularity  in the density of states at the gap edge.
The impurity band is initially concentrated around the energy of the single-impurity bound state, yet widens with increasing $\cimp$. 
It eventually merges with the continuum spectrum above the gap edge and fills the whole superconducting gap. This explains the phenomenon of gapless superconductivity\cite{AG61_eng}.

A separate class of disorder in superconductors is due to the inhomogeneities of the superconducting pairing potential, which can be induced, for instance, by random spatial variations of the coupling constant. Larkin and Ovchinnikov\cite{LO72JETP_disorder_eng} demonstrated the smearing of the BCS singularity by disorder of this type (see also Refs.~\onlinecite{Meyer+2001PRB,Skvortsov+2013JETP_subgap}). The shape of the smearing is essentially the same as for magnetic disorder and was argued to be universal\cite{Maki_Gapless,Tinkham_Introduction,Skvortsov+2013JETP_subgap} for all depairing mechanisms. The absence of impurity bands in Refs.~\onlinecite{LO72JETP_disorder_eng,Meyer+2001PRB,Skvortsov+2013JETP_subgap} at low disorder is a property of the model: here, the pairing-potential disorder was not associated with distinct impurities.  
A different situation, corresponding to pairing-potential impurities not
overlapping with other impurities, has been analyzed in Refs.~\onlinecite{Weinkauf+75ZPhys,Flatte+97PRB,Chattopadhyay+2002JPCM,Andersen+2006PRL}
(see also references therein for studies of $d$-wave superconductors).
According to Refs.~\onlinecite{Weinkauf+75ZPhys,Chattopadhyay+2002JPCM,Andersen+2006PRL}, a point-like impurity with suppressed pairing always supports a bound state. A numerical study of impurities with a size of the order of the Fermi wavelength $\lambda_F$\cite{Flatte+97PRB} did not find such a state when the ratio of the coherence length to $\lambda_F$ was sufficiently large. The formation of an impurity band at small impurity concentrations was discussed in Ref.~\onlinecite{Weinkauf+75ZPhys}. 

Generally, one may expect that, upon increasing the concentration of pair-breaking impurities, there is a complex crossover in the density of states near the gap edge. One starts from discrete impurity states below the gap edge. They form a narrow impurity band that widens and merges with the gap edge at some critical concentration. Upon further increasing the concentration, the complex shape of the density of states near the edge simplifies, approaching a universal one. The common potential scatterers do not influence this crossover, if the anisotropy of the pairing potential is neglected. However, in realistic situations, the anisotropy also modifies the density of states near the gap edge. 
In this paper, we present a detailed analysis of the crossover,
thus providing more insight into the properties of the bound quasiparticle states near the gap edge.

The impurity model we are mainly concerned with is a nonmagnetic scatterer that brings about a variation of the pair potential 
on a scale $L \ll \xi_S$, where $\xi_S$ is the coherence length in the pure limit. We 
evaluate the quasiclassical Green functions using the T-matrix approximation\cite{Balatsky+2006RMP}. We find that the  behavior of the density of states in the cases of pairing potential impurities and weak magnetic impurities is essentially the same, in a given spin sector, provided the latter are polarized along the same axis.  This analogy has strong implications, reproducing the sequence of the crossovers mentioned above. However, in contrast to magnetic impurities, the pair breaking impurities cannot completely close the superconducting gap at any realistic concentration. 

For localized impurity states,  we demonstrate that a spherically symmetric impurity with local suppression of
the order parameter gives rise to an infinite number of subgap bound states. We give explicit expressions for the energies $E_l$ of the states with orbital momentum $l$, and for the widths of the impurity bands at small impurity concentrations. The energy scale involved, $\Delta_0 - E_l$, is of the order of $\Delta_0 (L/\xi_S)^2 \ll \Delta_0$, where $\Delta_0$ is the bare superconducting order parameter.

It is almost forgotten nowadays that real superconductors have a slightly anisotropic gap, and with this  the density of states is sensitive to  common  potential disorder \cite{Anderson1959JPCS,Tsuneto62PTP,Hohenberg64JETP_eng,Clem66PR}. We derive the condition for the existence of  impurity states, which is modified by the anisotropy. A qualitative feature related to the anisotropy is a non-monotonic dependence of the gap-edge smearing on impurity concentration.  

The paper is organized as follows. In Sec.~\ref{sec:Basic} we introduce the model for the impurities and derive general equations within the T-matrix approximation for the quasiclassical Green functions. In Sec.~\ref{sec:isotropic} we analyze the case of isotropic pairing. In the limit of vanishing impurity concentration, we elucidate the properties of the impurity-bound states. At finite concentrations, we investigate and illustrate the crossover mentioned above. Section \ref{sec:anisotropic} considers the effects of a pairing anisotropy. We show that, in dirty superconductors, the remaining anisotropy  leads to a \lq\lq universal\rq\rq\ broadening of the gap edge. Thus, in general, the anisotropy affects the presence of impurity bound states, and we derive the condition for this. We analyze in detail the case of dirty superconductors at finite concentration of
potential impurities and illustrate this with plots.
We give our conclusions in Sec.~\ref{sec:Conclusion}. Several technical details are relegated to Appendices.

\section{General relations for the Green function and the T-matrix}
\label{sec:Basic}

A general disordered superconductor can be characterized by a Hamiltonian
\begin{equation}
  \hat{H} \! = \!\! \sum_{\alpha} \! \int \hat{\psi}^+_{\alpha}(\vec{r}) \!\! \left[ - \frac{\hbar^2}{2m} \frac{\partial^2}{\partial \vec{r}^2} - \mu \! + \!V(\vec{r}) \right] \! \hat{\psi}_{\alpha}(\vec{r}) d^3 \vec{r} + \! \hat{H}_{\mathrm{S}},
	\label{eq:H}
\end{equation}
where
\begin{eqnarray}
	 \hat{H}_{\mathrm{S}} &=& \int \Delta^*\left( \frac{\vec{p} + \vec{k}}{2}, \vec{p} - \vec{k} \right) \hat{\psi}_{\downarrow}(\vec{p}) \hat{\psi}_{\uparrow}(-\vec{k}) \frac{d^3\vec{p}}{(2\pi)^3} \frac{d^3 \vec{k}}{(2\pi)^3}  \nonumber \\
&&\,+\,\rm{h.c.}	
	\label{eq:H_SC}
\end{eqnarray}
describes electron pairing within mean-field theory.
Here $\hat{\psi}_{\alpha}(\vec{r})$ and $\hat{\psi}^+_{\alpha}(\vec{r})$ are the electron field operators, $\alpha = \{\uparrow,\downarrow\}$ is a spin label, $\mu$ is the chemical potential,  $V(\vec{r})$ is an electric impurity potential, and
\begin{equation}
	\hat{\psi}_{\alpha}(\vec{p}) = \!\!\! \int \!\!\hat{\psi}_{\alpha}(\vec{r}) e^{-i\vec{p}\vec{r}} d^3 \vec{r}, \quad \hat{\psi}^+_{\alpha}(\vec{p}) = \!\!\! \int \!\!\hat{\psi}^+_{\alpha}(\vec{r}) e^{i\vec{p}\vec{r}} d^3 \vec{r}.
	\label{eq:psi(p)}
\end{equation}
Note that the pairing potential $\Delta$ depends on two arguments, which reflect the pairing strength along the Fermi surface and its spatial variation, respectively. 

To determine the density of states associated with the Hamiltonian \eqref{eq:H}, we introduce the real-time retarded Green functions defined as
\begin{eqnarray}
	& G(\vec{r},\vec{r}',t) = -i \mean{\hat{\psi}_{\downarrow}(\vec{r},t) \hat{\psi}^+_{\downarrow} (\vec{r}',0) + \hat{\psi}^+_{\downarrow} (\vec{r}',0) \hat{\psi}_{\downarrow}(\vec{r},t)}, & \nonumber  \\ 
	& F(\vec{r},\vec{r}',t) = i \mean{\hat{\psi}_{\downarrow}(\vec{r},t) \hat{\psi}_{\uparrow} (\vec{r}',0) + \hat{\psi}_{\uparrow} (\vec{r}',0) \hat{\psi}_{\downarrow}(\vec{r},t)}, & \nonumber \\
	& F^+(\vec{r},\vec{r}',t) = i \mean{\hat{\psi}^+_{\uparrow}(\vec{r},t) \hat{\psi}_{\downarrow} (\vec{r}',0) + \hat{\psi}_{\downarrow} (\vec{r}',0) \hat{\psi}^+_{\uparrow}(\vec{r},t)}, & \nonumber \\
	& \hspace{-0.3cm} \bar{G}(\vec{r},\vec{r}',t) = i \mean{\hat{\psi}_{\uparrow}^+(\vec{r},t) \hat{\psi}_{\uparrow} (\vec{r}',0) + \hat{\psi}_{\uparrow} (\vec{r}',0) \hat{\psi}_{\uparrow}^+(\vec{r},t)} & \label{eq:FG}
\end{eqnarray}
at $t>0$ , and $G = F = F^+ = \bar{G} = 0$ at $t<0$. Here, the field operators $\hat\psi$ are in the Heisenberg representation. The Green functions satisfy the conventional Gor'kov equation, which in momentum representation reads
\begin{widetext}
\begin{eqnarray}
	 \left(
	\begin{array}{cc}
	  E + i\epsilon^+ \, - \xi(p) & 0 \\
		0 & -E - i\epsilon^+ - \xi(p)
	\end{array} \right)
	\check{G}_{E}(\vec{p},\vec{p}') && \nonumber \\
	 - \int \left(
	\begin{array}{cc}
	  V(\vec{p} - \vec{k}) & - \Delta\left(\frac{\vec{p} + \vec{k}}{2},\vec{p} - \vec{k} \right) \\
		\Delta^*\left(\frac{\vec{p} + \vec{k}}{2},\vec{k} - \vec{p} \right) & V(\vec{p} - \vec{k})
	\end{array} \right) \check{G}_{E}(\vec{k},\vec{p}') \frac{d^3 \vec{k}}{(2\pi)^3} &=&  (2\pi)^3\delta(\vec{p}-\vec{p}') \check{1},
	\label{eq:Gorkov0}
\end{eqnarray}
where $\epsilon^+$ is an infinitely small positive quantity, $\xi(p)$ is the kinetic energy measured from the Fermi level,
\begin{equation}
	\xi(p) = \frac{\hbar^2 p^2}{2m} - \mu = \frac{\hbar^2}{2m} (p^2 - k_F^2),
	\label{eq:xi}
\end{equation}
with the Fermi wave number $k_F = 2\pi/\lambda_F$, $V(\vec{p})$ is the Fourier transformed electric potential, and $\check{G}_{E}$ is a matrix composed of the Fourier transformed Green functions,
\begin{equation}
  \check{G}_{E}(\vec{p},\vec{p}') = \left(
	\begin{array}{cc}
	  G_{E} (\vec{p},\vec{p}') & F_{E} (\vec{p},\vec{p}') \\
		-F^+_{E}(\vec{p},\vec{p}') & \bar{G}_{E} (\vec{p},\vec{p}')
	\end{array}
	\right) 
	 = \!\! \int \! \left(
	\begin{array}{cc}
	  G(\vec{r},\vec{r}',t) & F(\vec{r},\vec{r}',t) \\
		-F^+ (\vec{r},\vec{r}',t) & \bar{G}(\vec{r},\vec{r}',t)
	\end{array} \right) \!
	e^{iE t/\hbar - i\vec{p} \vec{r} + i\vec{p}' \vec{r}'} {d^3 \vec{r} d^3 \vec{r}'}\frac{dt}{\hbar}.
	\label{eq:G_omega}
\end{equation}
\end{widetext}
In a clean superconductor, $V = 0$, the order parameter is spatially uniform, $\Delta(\vec{Q},\vec{q}) = (2\pi)^3 \Delta_0(\vec{Q}) \delta(\vec{q})$, and the translation invariance of the Green function yields $\check{G}_{E}(\vec{p},\vec{p}') =(2\pi)^3 \check{G}^{(0)}_{E}(\vec{p})\delta(\vec{p}-\vec{p}')$. Using Eq.~\eqref{eq:Gorkov0}, we obtain
\begin{equation}
	\Gnaught \! (\vec{p}) \! = \! \left(
	\begin{array}{cc}
	  \!\!E \!+\! i\epsilon^+  - \xi(\vec{p})\!\!\! & \Delta_0(\vec{p}) \\
		-\Delta_0^*(\vec{p}) &\!\!\! -E -\!i\epsilon^+ \!-\!\xi(\vec{p}) \!\!
	\end{array}
		\right)^{-1} \!\!.
	\label{eq:G0}
\end{equation}

For a start, let us assume that the disorder in the superconductor is induced by identical impurities with size $L$, whose positions are given by a set of vectors $\vec{R}_i$. Then, the pairing potential and the electric potential are 
\begin{eqnarray}
	\Delta(\vec{Q},\vec{q}) &=& (2\pi)^3 \Delta_0(\vec{Q}) \delta(\vec{q}) + \Delta_1 (\vec{Q},\vec{q}) \sum_{i} e^{-i\vec{q} \vec{R}_i},
	\label{eq:Delta_imp}
	\\
	V(\vec{q}) &=& U(\vec{q}) \sum_i e^{-i\vec{q} \vec{R}_i},
	\label{eq:U}
\end{eqnarray}
where the functions $\Delta_1 (\vec{Q},\vec{q})$ and $U(\vec{q})$ give the distortion of the pairing potential and the electric potential induced by a single impurity, respectively. We will evaluate the Green functions averaged over impurity positions, $\langle\check{G}_{E}(\vec{p},\vec{p}')\rangle_{\rm av}$, assuming a homogeneous distribution of the impurities. Then, the averaging procedure restores translational invariance, so that $\langle\check{G}_{E}(\vec{p},\vec{p}')\rangle_{\rm av}=  (2\pi)^3 \check{G}_{E}(\vec{p})\delta(\vec{p} - \vec{p}')$. 

Usually the impurity potential is taken into account in the second-order Born approximation \cite{Kopnin-book}. For our purposes this is not sufficient, since this approach does not yield localized impurity states. Instead, we make use of the more general T-matrix approximation (see Ref.~\onlinecite{Balatsky+2006RMP}, for example), which accounts for multiple scattering off each impurity. Within this appoximation, we will derive an equation for the quasiclassical Green functions. 

The T-matrix and the Green functions are determined from the following system of equations:
\begin{eqnarray}
	\check{T}_E(\vec{p},\vec{p}') &=& \Vimp(\vec{p},\vec{p}')
	\label{eq:T}\\
	&& \!+\! \int \!\Vimp(\vec{p},\vec{k}) \check{G}_{E}(\vec{k}) \check{T}_E(\vec{k},\vec{p}') \frac{d^3 \vec{k}}{(2\pi)^3},
	\nonumber
	\\
	\check{G}_{E}(\vec{p}) &=& \left[ \Gnaught(\vec{p})^{-1} - \cimp \check{T}_E(\vec{p},\vec{p}) \right]^{-1},
	\label{eq:main}
\end{eqnarray}
where
\begin{equation}
	\Vimp(\vec{p},\vec{k}) = \left(
  \begin{array}{cc}
	  U(\vec{p}-\vec{k}) & -\Delta_1\left(\frac{\vec{p} + \vec{k}}{2},\vec{p} - \vec{k} \right) \\
		\Delta_1^*\left(\frac{\vec{p} + \vec{k}}{2},\vec{k} - \vec{p} \right) & U(\vec{p}-\vec{k})
	\end{array}
	\right).
	\label{eq:Vimp}
\end{equation}
%
Equations \eqref{eq:T} and \eqref{eq:main} can be simplified for momenta close to the Fermi surface. Assuming that the Fermi energy is the largest energy scale, let us introduce the quasiclassical Green functions,
\begin{equation}
	\check{g} (E,\vec{n}) = \frac i\pi \int \check{G}_{E}(p\vec{n})\, d\xi(p),
	\label{eq:principal}
\end{equation}
where $\vec{n}$ is a unit vector, and integration is performed over a relatively small energy range, $\abs{\xi(p)} \ll \mu$. For simplicity, we restrict ourselves to the case of real functions $\Delta_0(\vec{Q})$ and $\Delta_1(\vec{Q},\vec{q})$ (a phase shift between $\Delta_0$ and $\Delta_1$ would manifest the violation of time-reversal symmetry). Then, the matrix $\check{g}$ has only two independent components,
\begin{equation}
	\check{g}(E,\vec{n}) = \left(
	\begin{array}{cc}
	  g_1(E,\vec{n}) & g_2(E,\vec{n}) \\
		-g_2(E,\vec{n}) & -g_1(E,\vec{n})
	\end{array} \right).
	\label{eq:g12}
\end{equation}
The density of states per spin 
is given by
\begin{equation}
	\nu(E) = \nu_0 \int \Re[g_1 (E,\vec{n})] \frac{d\vec{n}}{4\pi},
	\label{eq:nu}
\end{equation}
where $\nu_0 = k_F^3/(4\pi^2 \mu)$ is the density of states at the Fermi surface in the normal state for one spin direction.
Under the assumptions that the dependence of $\Delta_0(p\vec{n})$ and $\check{T}_E(p\vec{n},p\vec{n})$  on $p$ may be neglected when $p$ is close to $k_F$, it can be proved (see Appendix \ref{app:T}) that the matrix $\check{g}$ satisfies the relations
\begin{equation}
	\check{g}(E,\vec{n}) \check{S}_E(\vec{n}) - \check{S}_E(\vec{n}) \check{g}(E,\vec{n}) = 0
	\label{eq:commute}
\end{equation}
and
\begin{equation}
	g_1^2(E,\vec{n}) - g_2^2(E,\vec{n}) = 1,
	\label{eq:det}
\end{equation}
where
\begin{equation}
	\check{S}_E(\vec{n}) = \left(
	\begin{array}{cc}
	  E + i\epsilon^+ & \Delta_0(\vec{n}) \\
		-\Delta_0(\vec{n}) & -E - i\epsilon^+
	\end{array} \right)
	-\frac{\cimp}{\pi \nu_0} \check{T}_E(\vec{n},\vec{n})
	\label{eq:S}
\end{equation}
with $\Delta_0(\vec{n}) \equiv \Delta_0(k_F \vec{n})$, and $\check{T}_E(\vec{n},\vec{n}') \equiv \pi \nu_0 \check{T}_E(k_F\vec{n},k_F\vec{n}')$. 
Actually, Eq.~\eqref{eq:commute} is the standard Eilenberger equation for a macroscopically homogeneous superconductor \cite{Kopnin-book}. Equation \eqref{eq:det} expresses the normalization condition $\check{g}^2 = \check{1}$ for the quasiclassical Green function in the Eilenberger equation.

We assume that the spatial range of the pairing potential distortion $\Delta_1$ and of the electric potential $U$ is much smaller than the coherence length in the clean limit, $\xi_S = \hbar v_F/\pi \Delta_0$, where $v_F = \hbar k_F/m$ is the Fermi velocity.
In this case, Eq.~\eqref{eq:T} can be further simplified. To do this, we introduce an auxiliary normal-state scattering matrix $\check{f}(\vec{n},\vec{n}')$ that satisfies the equation
\begin{equation}
	\check{f} (\vec{p},\vec{p}') = \Vimp(\vec{p},\vec{p}') + \int \check{f}(\vec{p},\vec{k}) \check{\calG}(k) \Vimp(\vec{k},\vec{p}') \frac{d^3 \vec{k}}{(2\pi)^3},
	\label{eq:f}
\end{equation}
where $\check{\cal G}(k)=\check{G}_E^{(0)}({\bf k})$ taken at $\Delta_0=0$ and $E = 0$. The diagonal components of $\check{f}(\vec{p},\vec{p}')$ have the meaning of the dimensionless electron and hole scattering amplitudes off an impurity in the normal state. The off-diagonal components are the amplitudes of Andreev reflection of electrons and holes. Within the quasiclassical approximation, Eq.~\eqref{eq:T} can then be rewritten as
 (see Appendix \ref{app:T})
\begin{eqnarray}
	\check{T}_E (\vec{n},\vec{n}')\! &=& \! \check{f}(\vec{n},\vec{n}')  \label{eq:T_main}\\
	&&+ i \!\! \int \!\!\! \check{f}(\vec{n},\vec{n}'') [\check{\tau}_z - \check{g}(E,\vec{n}'')] 
	\check{T}_E (\vec{n}'' \!\! ,\vec{n}') \frac{d \vec{n}''}{4\pi},
	\nonumber
\end{eqnarray}
where $\check{\tau}_z$ is the third Pauli matrix acting in Nambu space, and $\check{f}(\vec{n},\vec{n}')\equiv\pi\nu_0\check f(k_F\vec{n},k_F\vec{n}')$. In Appendix \ref{app:f} the matrix $\check{f}(\vec{n},\vec{n}')$ is calculated for a spherically symmetric impurity with
\begin{equation}
	\Vimp(\vec{p},\vec{k}) = \left(
  \begin{array}{cc}
	  U(\vec{p}-\vec{k}) & -\Delta_1 \left(\vec{p} - \vec{k} \right) \\
		\Delta_1 \left(\vec{p} - \vec{k} \right) & U(\vec{p}-\vec{k})
	\end{array}
	\right).
	\label{eq:Vimp_iso}
\end{equation}

In the next two sections, we will solve the equations for the matrices $\check{T}_E$ and $\check{g}$ and analyze the resulting density of states in the cases of an isotropic and weakly anisotropic gap $\Delta_0(\vec{n})$, respectively.

\section{Superconductor with an isotropic gap}
\label{sec:isotropic}

We start with the case of isotropic pairing, when $\Delta_0(\vec{n}) = \mathrm{const}$ and $\Delta_1(\vec{Q},\vec{q}) = \Delta_1(\vec{q})$. Without loss of generality, we may then choose $\Delta_0 > 0$. If, additionally, the impurities are spherically symmetric, the matrix $\check{g}(E,\vec{n})$ will not depend on $\vec{n}$, and the matrix $\check{f}$ will have the form
\begin{equation}
	\check{f}(\vec{n},\vec{n}') = \left(
	\begin{array}{cc}
	  f_1(\vec{n},\vec{n}') &  f_2(\vec{n},\vec{n}') \\
		- f_2^*(\vec{n},\vec{n}') &  f_1^*(\vec{n},\vec{n}')
	\end{array} \right).
	\label{eq:f_components}
\end{equation}
To solve Eq.~\eqref{eq:T_main}, we expand $\check{f}$ and $\check{T}_E$ in terms of Legendre polynomials $P_l$:
\begin{eqnarray}
	 \check{T}_E(\vec{n},\vec{n}')  &=& \sum_{l=0}^{\infty} (2l+1) \check{T}_l(E) P_l(\vec{n} \cdot \vec{n}'),  \label{eq:T_expand} \\
	 \check{f}(\vec{n},\vec{n}') &=&  \sum_{l=0}^{\infty} (2l+1) \check{f}_l P_l(\vec{n} \cdot \vec{n}').  \label{eq:f_expand}
\end{eqnarray}
Using the addition theorem for spherical harmonics [Eq.~\eqref{eq:addition}], on can show that this leads to separate equations for the components $\check{T}_l$ with different orbital indices $l$. In particular, Eq.~\eqref{eq:T_main} yields  
\begin{equation}
	\check{T}_l(E) = \left\{ \check{1} - i \check{f}_l [\check{\tau}_z - \check{g}(E)] \right\}^{-1} \check{f}_l.
	\label{eq:Tl}
\end{equation}
To transform the right-hand side of this relation, it is convenient to use Eq.~\eqref{eq:Im(f_l)}, which is a corollary of a generalized optical theorem [Eq.~\eqref{eq:Im(f)}]. We obtain from Eqs.~\eqref{eq:T_expand} and \eqref{eq:Tl}
\begin{equation}
	\check{T}_E(\vec{n},\vec{n}) = \sum_{l=0}^{\infty} (2l+1) \frac{ \check{f}_l + i[\check{g}(E) -\check{\tau}_z] \Im[f_{1l}]}{1 - 2i f_{2l} g_2(E)}.
	\label{eq:T_simplified}
\end{equation}
Substituting Eq.~\eqref{eq:T_simplified} into Eq.~\eqref{eq:commute} yields
\begin{equation}
	E g_2(E) - \left[ \Delta_0 - \frac{\cimp}{\pi \nu_0} \sum_{l=0}^{\infty} \frac{(2l+1) f_{2l}}{ 1 - 2i f_{2l} g_2(E)} \right] g_1(E) = 0.
	\label{eq:g1g2_general}
\end{equation}
Thus, Eq.~\eqref{eq:g1g2_general} defines the Green functions in terms of the off-diagonal scattering amplitudes $f_{2l}$. Note that a very similar relation can be derived for weak polarized magnetic impurities, see Sec.~\ref{sub:Magnetic}. 

Near the gap edge, when $\abs{E - \Delta_0} \ll \Delta_0$, both $g_1$ and $g_2$ are large, $\abs{g_1},\abs{g_2} \gg 1$, and the normalization condition \eqref{eq:det} gives $g_2 \approx g_1 - 1/2g_{1}$. Thus, Eq.~\eqref{eq:g1g2_general} may be reduced to an equation for $g_1$ only. Namely,
\begin{equation}
	(E - \Delta_0) g_1 - \frac{\Delta_0}{2g_1} + \frac{\cimp}{\pi \nu_0} \sum_{l=0}^{\infty} \frac{(2l+1)f_{2l}}{ 1 - 2 if_{2l} g_1} g_1 = 0.
	\label{eq:edge_iso0}
\end{equation}

An explicit calculation of the coefficients $f_{2l}$ is given in Appendix \ref{app:f}. Under the assumption that $\abs{f_{2l}} \ll 1$ and for $l^2 \ll k_F \xi_S$, we find that these coefficients are given by
\begin{equation}
	f_{2l} = -\frac{\pi^2 \nu_0}{k_F^2} \int_0^{\infty}  \Delta_1(r) \abs{u_l(r)}^2dr.
	\label{eq:f12l}
\end{equation}
The functions $u_l(r)$, defined in Appendix \ref{app:f}, are the solutions of the Schr\"odinger equation in the normal state in the presence of the electric potential $U(r)$ only. 
If, furthermore, $ l+ 1/2\lesssim k_FL$, the amplitudes $f_{2l}$ can be estimated as 
\begin{equation}
	f_{2l} \sim \frac{\Delta_1}{\Delta_0}\frac{L}{\xi_S}.
	\label{eq:flS_estimate}
\end{equation}
Thus, the applicability condition of Eq.~\eqref{eq:f12l}, $\abs{f_{2l}} \ll 1$,  is satisfied in the realistic situation when $\abs{\Delta_1} \lesssim \Delta_0$.

We would like to point out that within our model, in full agreement with Anderson's theorem\cite{Anderson1959JPCS}, common potential impurities have no effect on the density of states, since $f_{2l} = 0$ for such impurities, and thus their T-matrix commutes with $\check{g}$. 
Hence, Eq.~\eqref{eq:edge_iso0} is not 
modified if the material is in the dirty limit with respect to potential disorder, i.e., $\Delta_0 \tau \ll \hbar$, where $\tau$ is the mean free time due to this disorder.

\subsection{Impurity states}
\label{eq:States_iso}
A defect with suppressed pairing, i.e., $\Delta_1(r)<0$, supports a set of localized Andreev states that are similar to the well-known Shiba states \cite{Yu1965,Shiba68PTP,Rusinov1969JETP} generated by magnetic impurities. For a point-like defect the existence of a single Andreev state has been predicted in Refs.~\onlinecite{Weinkauf+75ZPhys,Chattopadhyay+2002JPCM}. Gunsenheimer and Hahn \cite{Gunsenheimer+96PhysB} found multiple localized states for a sufficiently large pairing defect with $L \gg \lambda_F$. Here, we generalize these results, demonstrating that a defect with $\Delta_1 <0$, in fact, supports an \textit{infinite} number of Andreev states.  

To calculate the energies of the localized quasiparticle states, one has to determine the poles of the T-matrix at $\cimp \to 0$ (or, equivalently, solve the Bogoliubov-de Gennes equation, see Appendix \ref{app:BdG}). They are obtained from the equation
\begin{equation}
	 1 - 2i f_{2l} g_2(E) = 0,
	 	\label{eq:T_poles}
\end{equation}
where the function $g_2(E)$ is taken at $\cimp = 0$, i.e., $g_2(E)=-i\Delta_0/\sqrt{\Delta_0^2-E^2}$. Since we assume $\abs{f_{2l}} \ll 1$, the energies $E_l$ of the bound states lie close to the gap edge and are given by $E_l = \Delta_0 - {\cal E}_l$, where
\begin{equation}
	{\cal E}_l = 2f_{2l}^2 \Delta_0.
	\label{eq:El}
\end{equation}
As stated above, Eq.~\eqref{eq:El} is applicable for $l^2 \ll k_F \xi_S$.
However, this does not limit the number of bound states: as shown in Appendix \ref{app:BdG}, there are bound states at arbitrary large $l$. For typical impurity parameters $k_F L \sim 1$, $\abs{\Delta_1} \lesssim \Delta_0$, we have ${\cal E}_l \sim \Delta_0^3/\mu^2$ when $l$ is of the order of unity. At larger orbital momenta, the energies of the bound states quickly approach the gap edge with growing $l$.
 Explicit expressions for the quasiparticle energies in the particular case of a step-like function $\Delta_1(r)$ are derived in Appendix \ref{app:BdG}.


\subsection{Finite impurity concentration}
\label{eq:iso_band}

Above we showed that a single impurity produces bound states. At a finite impurity concentration, one expects these states to hybridize and form impurity bands that may merge with the continuum at a sufficiently high impurity concentration. We employ a standard simplifying assumption of pure $s$-wave scattering, neglecting all scattering amplitudes $f_{2l}$, except $f_{20}$. We do this to restrict ourselves to a single bound state, avoiding the consideration of a complex series of bound states, corresponding to higher orbital momenta. The assumption of pure $s$-wave scattering is justified if $k_F L \sim 1$, so that $f_{20} \lesssim \Delta_1/\mu$. 
Then, it is convenient to characterize the impurity concentration by the maximum scattering rate
\begin{equation}
	\frac{1}{\tau_{\mathrm{u}}} = \frac{2\cimp}{\hbar \pi \nu_0},
	\label{eq:tau_min}
\end{equation}
produced by these impurities in the unitary limit. To reduce the number of parameters in Eq.~\eqref{eq:edge_iso0}, we rescale the Green functions as well as energy and impurity concentrations, introducing the following dimensionless quantities:
\begin{equation}
	G_1 = \sqrt{\frac{2 {\cal E}_0}{\Delta_0}} g_1, \quad \Omega = \frac{E - \Delta_0}{{\cal E}_0},
\quad
	P = \frac{\hbar \sqrt{2}}{4 \sqrt{{\cal E}_0 \Delta_0}\, \tau_{\mathrm{u}} }.
	\label{eq:P_iso}
\end{equation}
It can be seen that the values $P \sim 1$ are achieved at $\hbar/\tau_{\mathrm{u}} \sim \sqrt{{\cal E}_0 \Delta_0}$.
%
%
In the new notations, Eq.~\eqref{eq:edge_iso0} takes the form
\begin{equation}
	\Omega G_1 - \frac{1}{G_1} \pm \frac{P G_1}{1 \mp i G_1} = 0,
	\label{eq:edge_iso_dimensionless}
\end{equation}
where one should take the upper sign for $f_{20}>0$ (corresponding to $\Delta_1<0$), and the lower sign for $f_{20}<0$ (corresponding to $\Delta_1<0$). 

\begin{figure}[htb]
	\centering
		\includegraphics[width = 0.49\linewidth]{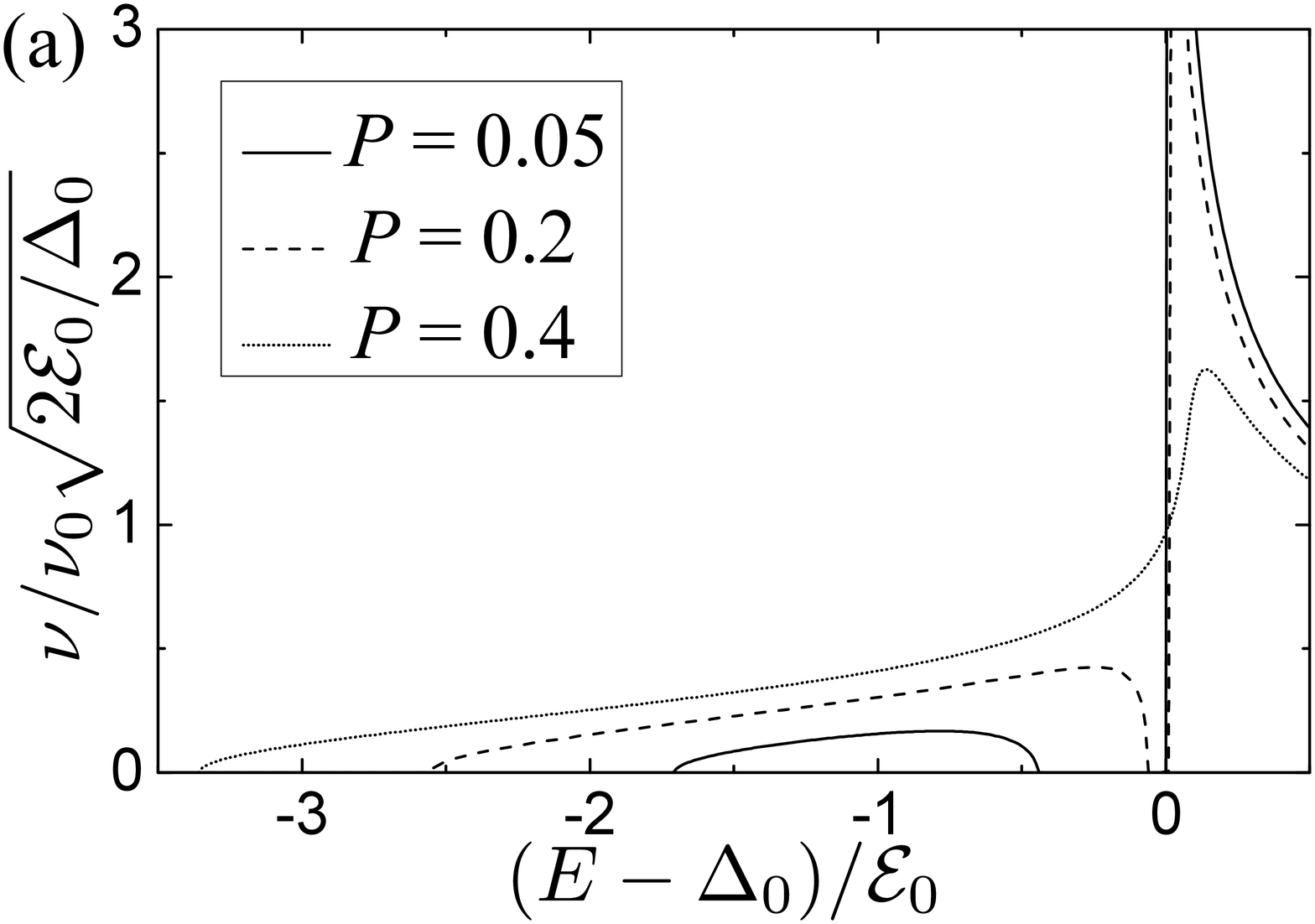}
		\includegraphics[width = 0.49\linewidth]{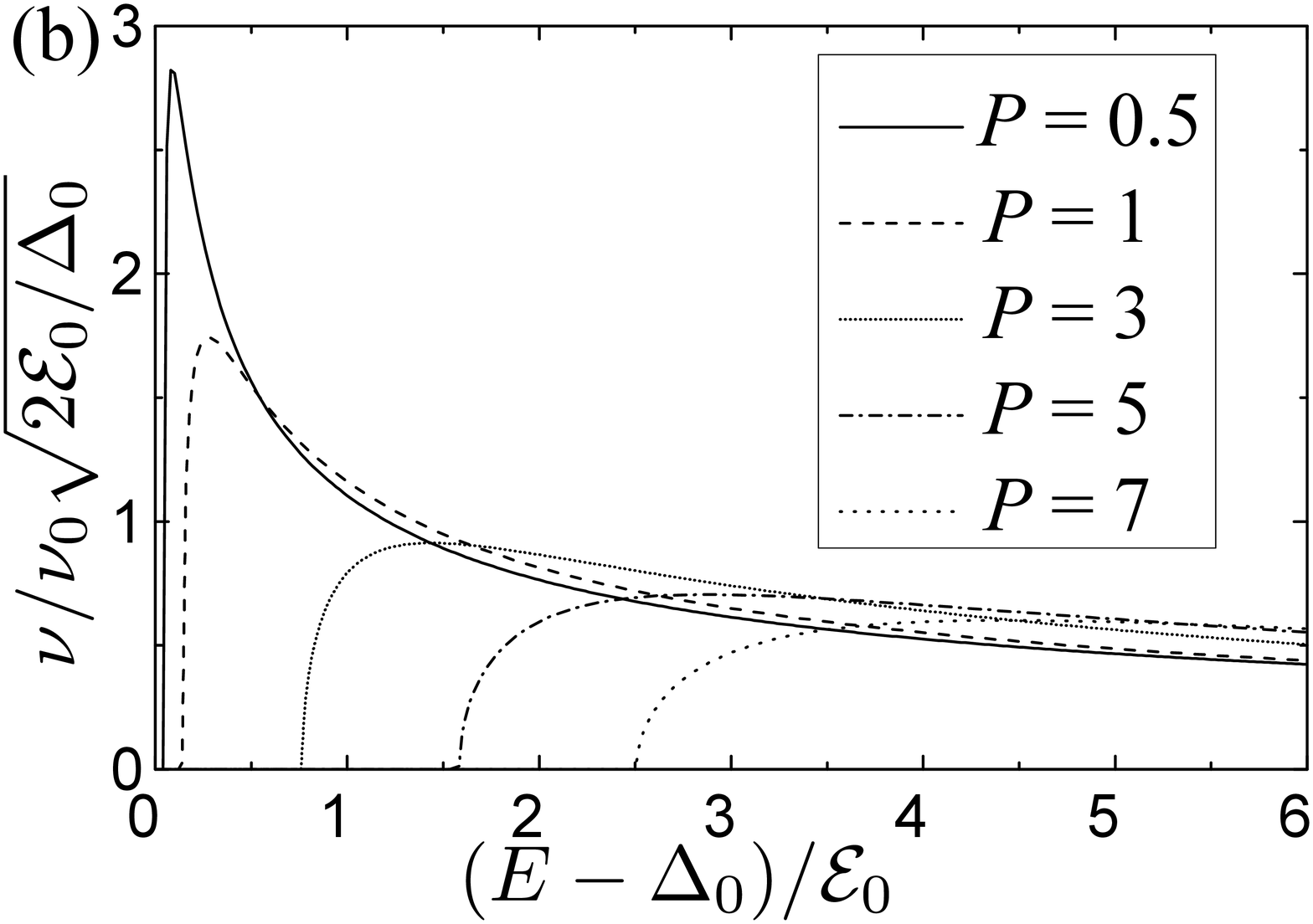}
	\caption{Energy dependence of the density of states in a superconductor with an isotropic gap containing pairing-potential impurities [Eq.~\eqref{eq:edge_iso_dimensionless}]. (a) $f_{20}>0$, (b) $f_{20}<0$.}
	\label{fig:Density_iso}
\end{figure}

When $f_{20}<0$, one finds a renormalized gap edge with a broadened BCS singularity. When $f_{20}>0$,  such that an individual impurity hosts bound states, the  localized states overlap at finite impurity concentration. At $0< P \ll 1$, they form a band centered around $\Omega  = -1$  with a width 
\begin{equation}
	W = 4 \sqrt{2P}.
	\label{eq:W_iso}
\end{equation}
Upon further increasing the impurity concentration, at $P = 8/27$ the impurity band merges with the continuum. The change of the energy dependence of the density of states with increasing $P$ is illustrated in Fig.~\ref{fig:Density_iso}.
When $P \gg 1$ the absolute value of $G_1$ becomes small at all values of $\Omega$, and Eq.~\eqref{eq:edge_iso_dimensionless} takes the form
\begin{equation}
	(\Omega \pm P) G_1 - \frac{1}{G_1} + iP G_1^2 = 0.
	\label{eq:edge_iso_universal}
\end{equation}
This equation describes  the behavior of the Green function in superconductors with pair-breaking impurities of different nature in the so-called universal limit, i.e., at sufficiently large impurity concentrations\cite{Maki_Gapless,Tinkham_Introduction,Skvortsov+2013JETP_subgap}. In this limit, the smoothing of the BCS singularity is commonly characterized by an effective depairing rate\cite{Skvortsov+2013JETP_subgap}, which equals in our case
\begin{equation}
	\frac{1}{\tdep} = \frac{f^2_{20}}{\tau_{\mathrm{u}}}.
	\label{eq:tdep_iso}
\end{equation}
It follows from Eq.~\eqref{eq:edge_iso_universal} that the gap edge is smeared on a scale of the order of
\begin{equation}
	\delta \Omega \sim P^{2/3},
	\label{eq:semaring_iso}
\end{equation}
and there is an additional shift of the gap edge by $\mp P$ due to the average pairing suppression/enhancement by the impurities. The characteristic values of the Green fuction are of the order of
$	G_1 \sim P^{-1/3}$.
%
These observations allow to rewrite Eq.~\eqref{eq:edge_iso_universal} in a form containing no parameters. Namely,
\begin{equation}
	\Omega' G_1' - \frac{1}{G_1'} + i G_1'^2 = 0,
	\label{eq:edge_one parameter}
\end{equation}
where $\Omega' = (\Omega \pm P) P^{-2/3}$ and $G_1' = G_1 P^{1/3}$. Note that the universal limit is approached rather slowly: corrections to $G_1'$ are of the order $P^{-1/3}$.

Using Eq.~\eqref{eq:edge_iso0}, now taking into account all components $f_{2l}$, we arrive at the following equation for $g_1$ in the universal limit:
\begin{equation}
	\left[ E - \Delta_0 + \frac{\hbar}{2\tau_{\mathrm{min}}} \sum_l (2l+1) f_{2l} \right] g_1  - \frac{\Delta_0}{2g_1} + i\frac{\hbar}{\tdep} g_1^2 = 0,
	\label{eq:universal_full}
\end{equation}
where the depairing rate equals
\begin{equation}
  \frac{1}{\tdep} = \frac{1}{\tau_{\mathrm{min}}} \sum_{l=0}^{\infty} (2l+1) (f_{2l})^2 = \frac{1}{\tau_{\mathrm{min}}} \int f_{2}^2(\vec{n},\vec{n}') \frac{d \vec{n}'}{4\pi}.
	\label{eq:tdep_full}
\end{equation}
Let us point out that Eq.~\eqref{eq:universal_full} was also obtained in the seminal paper by Larkin and Ovchinnikov\cite{LO72JETP_disorder_eng} for a different model of pairing-potential disorder. Namely, they assumed that the coupling constant exhibits fluctuations. Then, on the basis of the distribution of the coupling constants, the distortion of the pairing potential was calculated. This implies that there are fluctuations of the pairing potential on a scale that exceeds $\xi_S$. On the mean field level, this leads to a smoothing of the gap edge with a universal shape described by Eq.~\eqref{eq:universal_full}.
They evaluated the depairing rate for a superconductor with an arbitrary mean free path -- see Eq.~(21) in Ref.~\onlinecite{LO72JETP_disorder_eng}. For an infinite mean free path and $L \ll \xi_S$, that equation yields
\begin{equation}
	\left(\frac{1}{\tdep} \right)_{\rm LO} = \frac{\pi \nu_0 \cimp}{\hbar k_F^2} \int \int \frac{\Delta_1(\vec{r}) \Delta_1(\vec{r}')}{\abs{\vec{r} - \vec{r}'}^2} d^3 \vec{r} \,d^3 \vec{r}'.
	\label{eq:P'_quasiclassical}
\end{equation}
Within our model, we find the same result in the quasiclassical limit ($L \gg k_F^{-1}$) and for $U = 0$, when one should substitute $f_{2}(\vec{n},\vec{n}') = \pi \nu_0 \Delta_1 ( k_F(\vec{n} -\vec{n}'))$ in Eq.~\eqref{eq:tdep_full}.

\subsection{Comparison with magnetic impurities}
\label{sub:Magnetic}

Let us compare our results with the case of magnetic impurities that has been extensively studied in the literature. To describe such impurities, we add the following spin-dependent term to the Hamiltonian \eqref{eq:H}:
\begin{equation}
	\hat{H}_M = \sum_{\alpha,\beta} \int \hat{\psi}_{\alpha}^+ (\vec{r}) [\vec{J}(\vec{r}) \cdot\hat{\boldsymbol{\sigma}}]_{\alpha \beta}  \hat{\psi}_{\beta} (\vec{r}) d^3 \vec{r},
	\label{eq:H_M}
\end{equation}
where $\vec{J}(\vec{r})$ is the exchange field, and $\hat{\boldsymbol{\sigma}}$ is a vector composed of Pauli matrices acting in spin space. The exchange field is given by
\begin{equation}
	\vec{J}(\vec{r}) = \sum_{i} J_1(\vec{r} - \vec{R}_i) \vec{S}_i,
	\label{eq:J}
\end{equation}
where $J_1>0$, and the unit vectors $\vec{S}_i$ specify the polarizations of the impurities. 

Let us first assume that all impurities are polarized in the same direction, i.e., all vectors $\vec{S}_i$ are identical. When evaluating the Green functions, we may now neglect the distortion of the pairing potential induced by the impurities, since its effect on the density of states in realistic situations is much smaller than the influence of the exchange field \cite{Rusinov1969JETP}. Then, within the T-matrix approximation, we obtain the following relation [a similar calculation has been previously done in Ref.~\onlinecite{Fominov+2011PRB} for point-like impurities]:
\begin{equation}
	E g_2(E) -  \Delta_0 g_1(E) \pm \frac{\cimp}{\pi \nu_0} \sum_{l=0}^{\infty} \frac{(2l+1) f_l^M}{1 \mp 2 i f_l^M g_1(E)} g_2(E) = 0.
	\label{eq:g1g2_magnetic}
\end{equation}
Here, the upper/lower sign corresponds to \lq\lq spin-up\rq\rq/\lq\lq spin-down\rq\rq\ electrons with respect to the polarization direction, and $f_l^M$ are the differences of scattering amplitudes of \lq\lq spin-up\rq\rq\ and \lq\lq spin-down\rq\rq\ electrons. Under the constraints $\abs{f_l^M} \ll 1$ and $l^2 \ll k_F \xi_S$ one can prove that the magnetic coefficients $f_l^M$ are given by expressions similar to Eq.~\eqref{eq:f12l},
\begin{equation}
	f_l^M = -\frac{\pi^2 \nu_0}{k_F^2} \int_0^{\infty}J_1(r)  \abs{u_l(r)}^2 dr.
	\label{eq:flM}
\end{equation}
It can be seen that Eqs.~\eqref{eq:g1g2_magnetic} and \eqref{eq:g1g2_general} have almost the same form, the only difference being the permutation of $g_1$ and $g_2$ in the last term. 
However, this difference is not essential near the gap edge, when $\abs{E - \Delta_0} \ll \Delta_0$. As noted earlier, in that case $g_2 \approx g_1 - 1/2g_{1}$,  and Eq.~\eqref{eq:g1g2_magnetic} yields%
\begin{equation}
	(E - \Delta_0) g_1 \pm \frac{\cimp}{\pi \nu_0} \sum_{l=0}^{\infty} \frac{(2l+1) f_l^M}{1 \mp 2 i f_l^M g_1} g_1 = 0.
	\label{eq:edge_isoM}
\end{equation}
Comparing Eqs.~\eqref{eq:edge_iso0} and \eqref{eq:edge_isoM}, we see that pairing-potential impurities with $\Delta_1>0$ act like magnetic impurities in the \lq\lq spin-up\rq\rq\ sector whereas pairing-potential impurities with $\Delta_1<0$ act like magnetic impurities in the \lq\lq spin-down\rq\rq\ sector. The full density of states in the case of magnetic impurities is obtained by summing over both spin sectors.

\begin{figure}[htb]
	\centering
		\includegraphics[width = 0.49\linewidth]{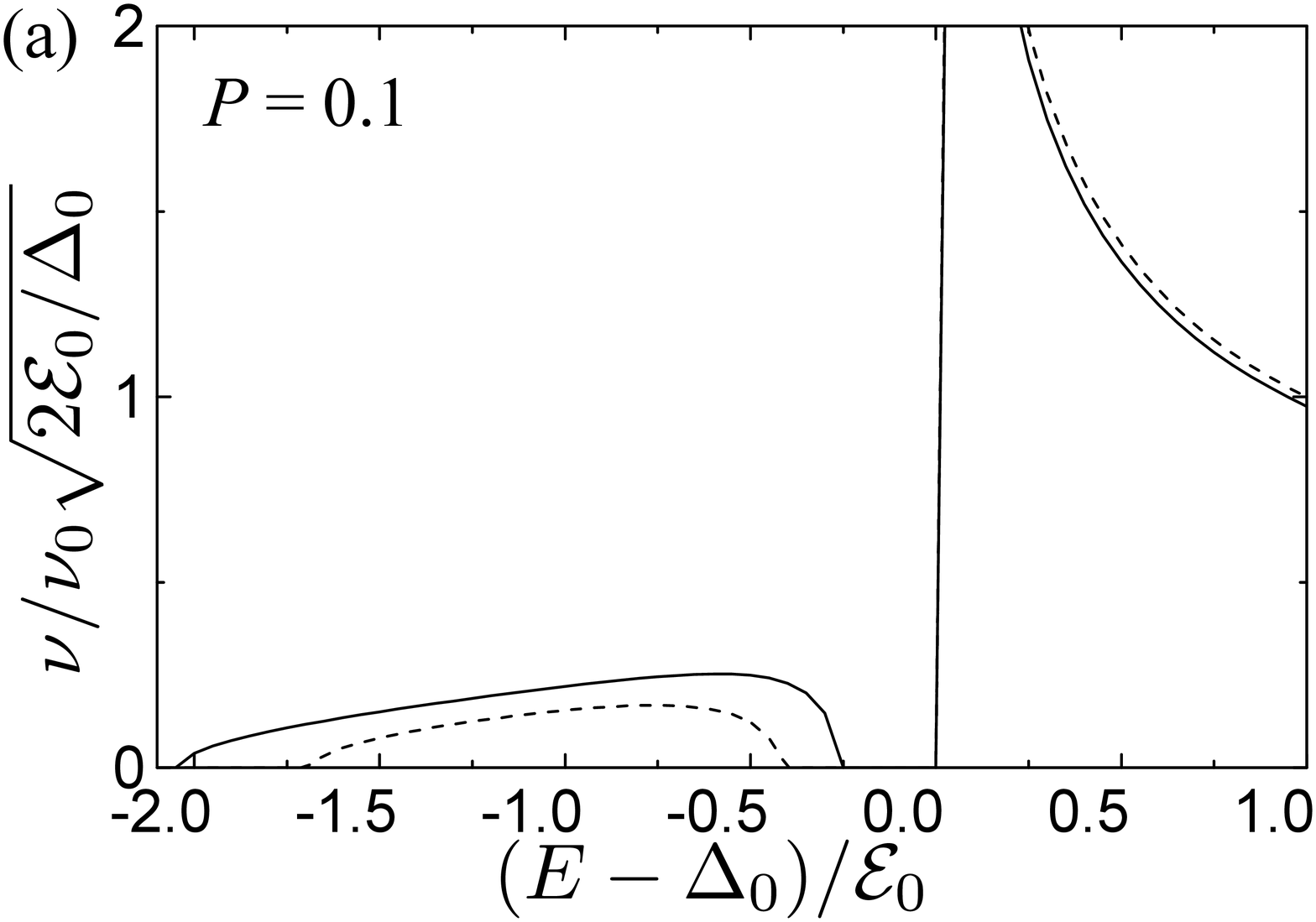}
		\includegraphics[width = 0.49\linewidth]{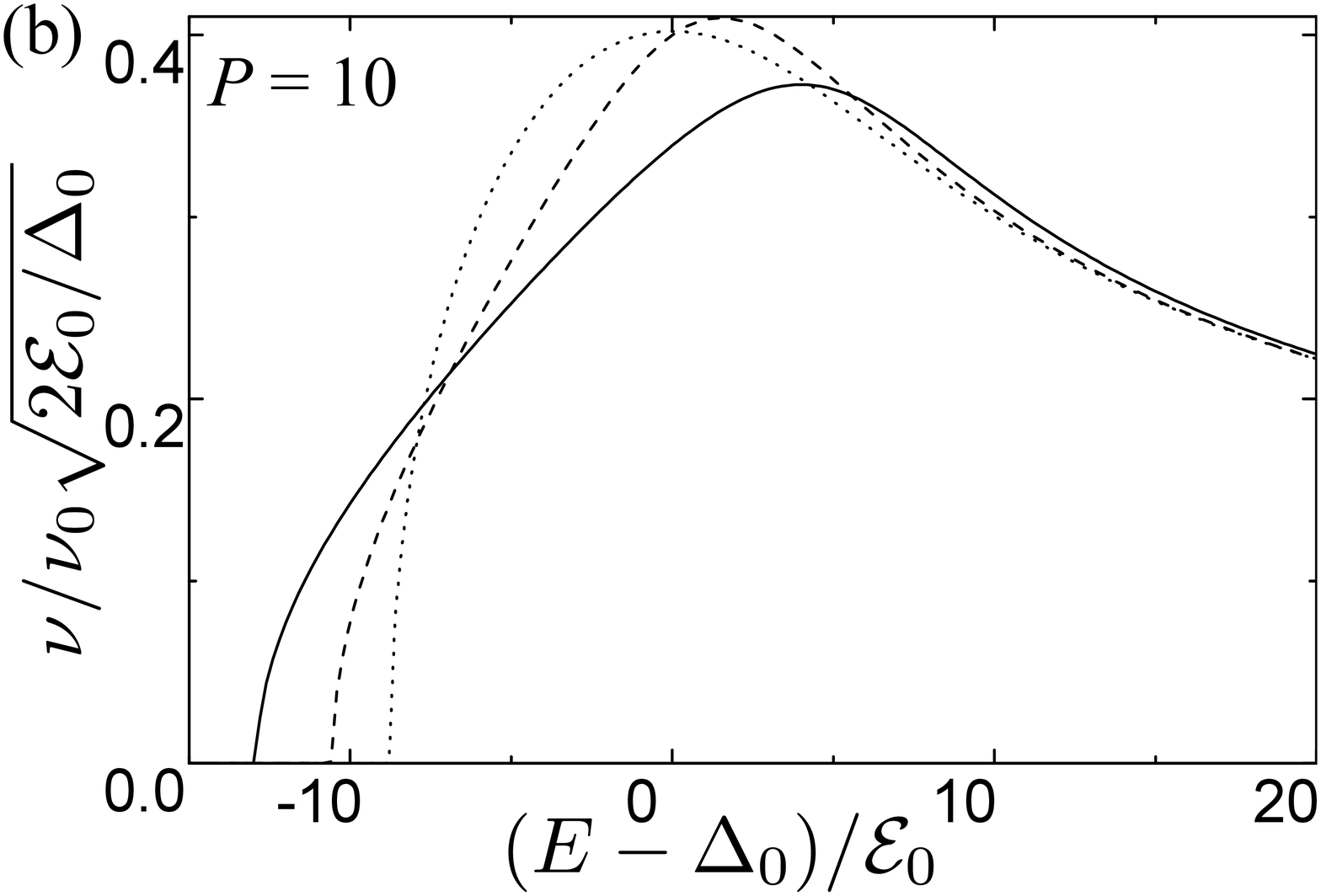}
	\caption{Densities of states vs. energy in a superconductor containing either impurities with a local gap [solid line, Eq.~\eqref{eq:edge_iso_dimensionless} with the upper signs] or magnetic impurities with randomly oriented spins [dashed line, Eq.~\eqref{eq:edge_unpolarized}], and the density of states in the universal limit [dotted line in graph (b), Eq.~\eqref{eq:edge_iso_universal}]. The additional shift $-P$ of the gap edge, appearing in Eqs.~\eqref{eq:edge_iso_dimensionless} and \eqref{eq:edge_iso_universal}, is compensated via a positive translation of the corresponding curves by $P{\cal E}_0$ in graph (b).}
	\label{fig:ShibaMine}
\end{figure}

The comparison with magnetic impurities can be extended to the case of \textit{randomly oriented} spins. To see this, we  recast the relation for the Green function derived in Ref.~\onlinecite{Rusinov1969JETP} to a form similar to Eq.~\eqref{eq:edge_iso_dimensionless},
\begin{equation}
	\Omega G_1 - \frac{1}{G_1} + \frac{i P G_1^2}{1 + G_1^2} = 0.
	\label{eq:edge_unpolarized}
\end{equation}
This equation is obtained by averaging Eq.~\eqref{eq:edge_iso_dimensionless} over impurity-spin directions, and accordingly $G_1$ is the Green function averaged over impurity-spin directions. At small $P$, Eq.~\eqref{eq:edge_unpolarized} gives an impurity band with a width $W = 4\sqrt{P}$. The merger of this band with the continuum occurs at $P \approx 0.49$. The plots shown in Fig.~\ref{fig:ShibaMine} demonstrate the qualitative similarity of the energy dependent densities of states derived from Eqs.~\eqref{eq:edge_iso_dimensionless} and \eqref{eq:edge_unpolarized}. 

When $P \gg 1$, Eq.~\eqref{eq:edge_unpolarized} reduces to the relation in the universal limit, Eq.~\eqref{eq:edge_one parameter}. Note than due to the averaging over spin directions, unlike in Eq. \eqref{eq:edge_iso_universal}, there is no additional gap shift $\mp P$. Moreover, the universal limit is approached faster than in the case of pairing-potntial impurities or polarized magnetic imputrities, as corrections to the Green function averaged over impurity-spin directions $G_1'$ are of the order $P^{-2/3}$ only.

Finally, let us mention that in the case of relatively strong magnetic impurities (with $J_1 \gg \Delta_0$) a gapless regime can be reached. By contrast, within the field of applicability of our approach, a superconductor with pairing-potential impurities is always gapped. Indeed, to reach the gapless regime we would need $\cimp f_{20}/\nu_0 \sim \Delta_0$, which requires $\cimp \gtrsim k_F^2 L^{-1}$ according to Eq.~\eqref{eq:flS_estimate}, i.e., at least $\cimp \gtrsim L^{-3}$. At such large concentrations, the impurities ``overlap'' and our simple model is not valid any longer. The estimates above also indicate that the quasiclassical Green functions and the density of states are modified only in narrow energy range, $\abs{E - \Delta_0} \ll \Delta_0$ at realistic impurity concentrations. As a consequence, in contrast to magnetic impurities, a self-consistent recalculation of the bulk pairing potential $\Delta_0$ is not required. 

\section{Superconductor with a weakly anisotropic gap}
\label{sec:anisotropic}

In any realistic superconductor the pairing potential is at least slightly anisotropic, i.e., $\Delta_0(\vec{n}) \neq \mathrm{const}$. The anisotropy used to be a subject of active theoretical and experimental research\cite{Geilikman1974Kinetic}, but has been largely ignored in modern models of $s$-wave superconductors. It is clear that even a small anisotropy significantly influences the spectral properties of the superconducting state in the vicinity of the gap edge. As such, it may modify the results of the previous section. In this section, we consider these modifications. 

We assume a weak anisotropy, so that the anisotropic part of the bulk pairing potential, $\Delta'(\vec{n}) \equiv \Delta_0(\vec{n})- \mean{\Delta_0}$, is small: $\abs{\Delta'} \ll \mean{\Delta_0}$. Here and further, the angle brackets denote the average over the Fermi surface,
\begin{equation}
	\mean{X} \equiv \int X(\vec{n}) \frac{d\vec{n}}{4\pi}.
	\label{eq:brackets}
\end{equation}
%
Under the assumption of a weak anisotropy of $\Delta_0(\vec{n})$, it is reasonable to neglect the anisotropy of the impurity pairing potential, i.e., put $\Delta_1(\vec{Q},\vec{q}) = \Delta_1(\vec{q})$. The matrix $\check{f}(\vec{n},\vec{n}')$ does not depend on $\Delta_0(\vec{n})$, hence, in the case of spherically symmetric impurities it is still given by Eq.~\eqref{eq:f_expand}, and the off-diagonal components of the expansion coefficients $f_{2l}$ are defined by Eq.~\eqref{eq:f12l}.
In the presence of anisotropy, the T-matrix is defined by Eq.~\eqref{eq:T_main}. Now, since the Green function $\check{g}(\omega,\vec{n})$ depends on $\vec{n}$, it is impractical to
solve this equation using the expansion in terms of Legendre polynomials.
To overcome this inconvenience, we employ again the approximation of $s$-wave scattering: $\check{f}(\vec{n},\vec{n}') = \check{f}_0 = \mathrm{const}$. In this case $\check{T}_E(\vec{n},\vec{n}')$ does not depend on $\vec{n}$ and $\vec{n}'$, and is given by
\begin{eqnarray}
	 \check{T}_E &=& [\check{f}^{-1} +i \mean{\check{g}} - i \check{\tau}_z]^{-1}  \nonumber \\
	& =& \frac{\Re[{\check{f}}] - i\mathrm{det} [\check{f}] \mean{\check{g}} }{1 - 2i f_{20} \mean{g_2} - \mathrm{det} [\check{f}] (\mathrm{det} \mean{\check{g}} + 1)}. \label{eq:T_aniso1}
\end{eqnarray}
We made use of Eq.~\eqref{eq:Im(f_l)} to arrive at the last line. Under the approximation of $s$-wave scattering, Eq.~\eqref{eq:T_aniso1} is valid even in the case of strong anisotropy.

As above, the energies of the impurity states correspond to the poles of the T-matrix, which are given by
\begin{equation}
	1 - 2i f_{20} \mean{g_2} - \mathrm{det} \check{f} (\mathrm{det} \mean{\check{g}} + 1) = 0.
	\label{eq:states_aniso}
\end{equation}
If the potential scattering is weak ($\abs{\mathrm{det} \check{f}} \ll 1$) or in the limit of sufficiently large impurity concentrations, when the Green functions are essentially isotropic, so that $\mathrm{det} \mean{\check{g}} \approx \mean{\mathrm{det} \, \check{g}} = -1$, we can neglect the third term in Eq.~\eqref{eq:states_aniso}.
%
Then, 
\begin{equation}
	\check{T}_E \approx \frac{\Re{[\check{f}}] -i \mathrm{det} [\check{f}]\mean{\check{g}} }{1 - 2i f_{2} \mean{g_2}}.
	\label{eq:T_aniso2}
\end{equation}
%

The third term in Eq.~\eqref{eq:states_aniso}, which is proportional to $\mathrm{det}[ \check{f}]$, generally, can not be neglected in a strongly anisotropic superconductor. In $d$-wave superconductors, this term is responsible for the quasibound states\cite{Balatsky+95PRB}, that possess a complex energy with a small imaginary part. Such states are absent in the case of weak anisotropy under consideration here.

\subsection{Ordinary potential scatterers}
\label{sub:ordinary_aniso}

Before addressing the influence of pairing-potential impurities on the density of states, we will briefly consider the effect of ordinary potential scatterers\cite{Tsuneto62PTP,Clem66PR}. 

At $\Delta_1 = 0$, one obtains $f_{20} = 0$, and Eq.~\eqref{eq:T_aniso2} reduces to
\begin{equation}
  \check{T}_E = \Re{\check{f}} -  i \abs{f_{10}}^2 \mean{\check{g}}.
	\label{eq:T_ordinary}
\end{equation}
The T-matrix  given by Eq.~\eqref{eq:T_ordinary} has no poles, so there are no subgap states. 

%
Equations \eqref{eq:commute} and \eqref{eq:det} yield
\begin{equation}
	g_1 =  - \frac{i\tilde{E}}{\sqrt{\tilde{\Delta}^2(\vec{n}) - \tilde{E}^2}}, \qquad g_2 = - \frac{i\tilde{\Delta}(\vec{n})}{\sqrt{\tilde{\Delta}^2(\vec{n}) - \tilde{E}^2}},
	\label{eq:g1g2_dimensionless}
\end{equation}
where
\begin{equation}
	\tilde{E} = E + \frac{i\hbar}{2\tau} \mean{g_1}, \quad \tilde{\Delta}(\vec{n}) = \Delta_0(\vec{n}) + \frac{i\hbar}{2\tau} \mean{g_2},
	\label{eq:with_Pimp}
\end{equation}
and the scattering time is given by
\begin{equation}
	\frac{1}{\tau} = \frac{\abs{f_{10}}^2}{\tau_{\mathrm{u}}}.
	\label{eq:tau}
\end{equation}
%
This is equivalent to a set of equations derived in Refs.~\onlinecite{Tsuneto62PTP} and \onlinecite{Clem66PR}.

In the case of weak anisotropy the density of states is affected by the scatterers only at energies near the gap edge, $\abs{E - \mean{\Delta_0}} \ll \mean{\Delta_0}$. In this energy range we can utilize that $\abs{\delta g} \ll \abs{\mean{g_1}}$, where $\delta g = \mean{g_1} - \mean{g_2}$. As a consequence, $\abs{\tilde{E} - \tilde{\Delta}(\vec{n})} \ll \abs{\tilde{E}}$, and
%
\begin{equation}
	\mean{g_1} \approx \frac{1}{\sqrt{2}} \mean{ \left( \frac{\tilde{E} - \tilde{\Delta}(\vec{n})}{\tilde{E}} \right)^{-\frac{1}{2}}}, 
	\label{eq:g1_edge0}
\end{equation}
\begin{equation}
	\delta g \approx \frac{1}{\sqrt{2}} \mean{ \left( \frac{\tilde{E} - \tilde{\Delta}(\vec{n})}{\tilde{E}} \right)^{\frac{1}{2}}},
	\label{eq:difference_edge0}
\end{equation}
or
\begin{equation}
	\mean{g_1} \! = \! \frac{\sqrt{\mean{\Delta_0} +\frac{i\hbar}{2\tau} \mean{g_1}}}{\sqrt{2}} \mean{ \left[ E - \Delta_0(\vec{n}) + \frac{i\hbar}{2\tau} \delta g \right]^{-\frac{1}{2}}} \!, 
	\label{eq:g1_edge}
\end{equation}
\begin{equation}
	\delta g = \frac{1}{\sqrt{2}\sqrt{\mean{\Delta_0} +\frac{i\hbar}{2\tau} \mean{g_1}}} \mean{ \left[ E - \Delta_0(\vec{n}) + \frac{i\hbar}{2\tau} \delta g \right]^{\frac{1}{2}}}.
	\label{eq:difference_edge}
\end{equation}
To understand the scaling of the Green function in the case of weak anisotropy, it is instructive to rewrite these relations in terms of the dimensionless quantities 
\begin{eqnarray}
	&   \delta(\vec{n})  = \frac{\Delta'(\vec{n})}{\sqrt{\mean{\Delta'^2}}}, \quad \tilde{G}_1 = \mean{g_1}\frac{ \sqrt{2} \sqrt[4]{\mean{\Delta'^2}}}{\sqrt{\mean{\Delta_0}}}, \quad \delta \tilde{G} = \delta g\frac{ \sqrt{2 \mean{\Delta_0}}}{\sqrt[4]{\mean{\Delta'^2}}}, & \nonumber \\
	& \tilde{P} = \frac{\hbar }{2\sqrt{2 \mean{\Delta_0}} \sqrt[4]{\mean{\Delta'^2}} \tau}, \quad \tilde{\Omega} = \frac{E - \mean{\Delta_0}}{\sqrt{\mean{\Delta'^2}}}. &
	\label{eq:everything_scaled}
\end{eqnarray}
Then we have
\begin{equation}
	\tilde{G}_1 \! = \! \sqrt{1 +i \tilde{P} \tilde{G}_1} \mean{ \left[ \tilde{\Omega} - \delta(\vec{n}) + i \tilde{P} \delta \tilde{G} \right]^{-\frac{1}{2}}} \!, 
	\label{eq:g1_edge'}
\end{equation}
\begin{equation}
	\delta \tilde{G} = \frac{1}{\sqrt{1 +i \tilde{P} \tilde{G}_1}} \mean{ \left[ \tilde{\Omega} - \delta(\vec{n}) + i \tilde{P} \delta \tilde{G} \right]^{\frac{1}{2}}}.
	\label{eq:difference_edge'}
\end{equation}
In the limit $\tilde{P} \ll 1$ of low concentrations of the scatterers, the gap edge is rounded at an energy scale $ \sqrt{\mean{\Delta'^2}}$. In the opposite limit of large impurity concentrations, the rounding becomes more narrow. This is due to the suppression of the anisotropy of the pairing potential, which becomes essential when $\tilde{P} \gg 1$. This condition is satisfied even in relatively clean superconductors, at $\mean{\Delta_0} \tau \gtrsim \hbar$.

Let us consider the limit of strong suppression of the anisotropy, $\tilde{P} \gg 1$. In this limit we can expand the expressions in the anglular brackets in Eqs.~\eqref{eq:g1_edge'} and \eqref{eq:difference_edge'} in terms of the small ratio $\delta(\vec{n})/(\tilde{P} \delta \tilde{G})$. 
Then, we can eliminate $\delta \tilde{G}$ from the equations to arrive at the relation
%
%
%
\begin{equation}
	\tilde{\Omega} \tilde{G}_1 - \frac{1}{\tilde{G}_1} + \frac{i}{2\tilde{P}} \tilde{G}_1^2 = 0.
	\label{eq:universal_aniso}
\end{equation}
We observe that Eq.~\eqref{eq:universal_aniso} is equivalent to the ``universal limit'' equation of Ref.~\onlinecite{Skvortsov+2013JETP_subgap}. Thus, the rounding of the gap edge owing to the weak anisotropy and potential scattering can also be described in the framework of this universal scheme. In this limit the density of states is isotropic in the main order, not depending on the details of the shape of $\Delta_0(\vec{n})$. The ``depairing rate'', as defined in Ref.~\onlinecite{Skvortsov+2013JETP_subgap}, equals
\begin{equation}
	\frac{1}{\tau_{\mathrm{dep}}} 
	= \frac{2\mean{\Delta'^2} \tau}{\hbar^2}.
	\label{eq:tau_dep}
\end{equation}
Finally, the substitutions 
\begin{equation}
	\tilde{G}_1 = G_1' (2\tilde{P})^{1/3}, \qquad \tilde{\Omega} = \frac{\Omega'}{(2\tilde{P})^{2/3}}.
	\label{eq:to_one_parameter}
\end{equation}
reduce Eq.~\eqref{eq:universal_aniso} to the form \eqref{eq:edge_one parameter}, containing no parameters. 
Note that the dependence $1/\tau_{\mathrm{dep}}\propto\tau$ in Eq.~\eqref{eq:tau_dep} reflects the gap edge sharpening with growing impurity concentration, which was discussed above.

\subsection{Suppressed anisotropy and pairing potential impurities}
\label{sub:2impurities}

Now we will analyze the situation when the material contains both common potential scatterers 
and pairing-potential impurities. 
The ordinary scatterers and pairing-potential impurities have the concentrations $c_1$ and $c_2$, and scattering amplitude matrices $\check{f}^{(1)}$ and $\check{f}^{(2)}$, respectively. The corresponding T-matrices are
\begin{equation}
	\check{T}_E^{(1)} = \Re{\check{f}^{(1)}} -  i \abs{f_{10}^{(1)}}^2 \mean{\check{g}}
	\label{eq:T_1}
\end{equation}
for ordinary scatterers and
\begin{equation}
	\check{T}_E^{(2)} = \frac{\Re{\check{f}^{(2)}} -  i \abs{f_{10}^{(2)}}^2 \mean{\check{g}}}{1 - 2i f_{20} \mean{g_2}}
	\label{eq:T_2}
\end{equation}
for pairing-potential impurities, where $f_{20} \equiv f_{20}^{(2)}$, similar to Sec.~\ref{sec:isotropic}. 
To determine the Green functions, we substitute in Eq.~\eqref{eq:commute} $\check{S}(E,\vec{n})$ in the form
\begin{equation}
	\check{S}(E,\vec{n}) = \left(
	\begin{array}{cc}
	  E+ i \epsilon^+ & \Delta_0(\vec{n}) \\
		-\Delta_0(\vec{n}) & -E - i\epsilon^+
	\end{array} \right)
	-\sum_{i=1,2} \frac{c_i}{\pi \nu_0} \check{T}_E^{(i)}(\vec{n},\vec{n}).
	\label{eq:S2}
\end{equation}
Then, $g_1(\vec{n})$ and $g_2(\vec{n})$ are given by Eq.~\eqref{eq:g1g2_dimensionless} with
\begin{eqnarray}
	& \tilde{E} = E + \frac{i\hbar}{2\tau(E)} \mean{g_1}, &
	\label{eq:E_tilde} \\
	 &\tilde{\Delta}(\vec{n}) = \mean{\Delta_0} (\Xi - Q) + \Delta'(\vec{n}), &
	\label{eq:Delta_tilde} \\
	& \frac{1}{\tau(E)} = \frac{1}{\tau_1} + \frac{1}{\tau_2(1- 2if_{20} \mean{g_2})}, \quad \frac{1}{\tau_{1,2}} = \frac{2c_{1,2}\abs{f_{10}^{(1,2)}}^2}{\hbar \pi \nu_0},& \label{eq:tau(E)} \\
	& \Xi = 1 + \frac{i\hbar}{2\tau(E) \mean{\Delta_0}} \mean{g_2}, & \label{eq:Xi} \\
	& Q = \frac{c_2 f_{20}}{\pi \nu_0 \mean{\Delta_0}} \frac{1}{1 - 2i f_{20} \mean{g_2}}, & \label{eq:Q}
\end{eqnarray}
$1/\tau_1$ and $1/\tau_2$ being the potential scattering rates due to ordinary scatterers and pairing-potential impurities, respectively. From Eq.~\eqref{eq:tau(E)} we can see that the contribution of pairing-potential impurities to potential scattering is enhanced at energies close to the energy of the bound state [see Eq.~\eqref{eq:T_poles}], manifesting resonant scattering near this energy. 

Let us now derive simplified equations, applicable in the vicinity of the gap edge ($\abs{E - \mean{\Delta_0}} \ll \mean{\Delta_0}$). To do this, let us notice that the quantities $\Xi$ and $-Q$ in Eq.~\eqref{eq:Delta_tilde} 
represent the renormalizations of the isotropic part of $\Delta_0(\vec{n})$ due to common superconducting and potential scattering, respectively. Simplifications are possible, if $Q$ is small. If the second fraction in Eq.~\eqref{eq:Q} is of the order of or smaller than unity, this statment is rather obvious since $\abs{Q} \lesssim c_2 \abs{f_{20}}/(\pi \nu_0 \mean{\Delta_0}) \lesssim c_2 k_F^{-3} \ll 1$, as estimated in Sec.~\ref{eq:iso_band}. The danger is that the second fraction in Eq.~\eqref{eq:Q} can become large close to the energy of the bound state.
Hovewer, at finite concentrations of the pairing-potential impurities, the largest value of this fraction is proportional to $1/\sqrt{c_2}$. 
Hence, $Q \propto \sqrt{c_2}$, and it vanishes at $c_2 \to 0$. This proves that $\abs{Q} \ll 1$. In turn, the smallness of $Q$ provides the validity of Eqs.~\eqref{eq:g1_edge0} and \eqref{eq:difference_edge0} with
%
%
\begin{equation}
	\frac{\tilde{E} - \tilde{\Delta}(\vec{n})}{\tilde{E}} = \frac{E - \Delta_0(\vec{n}) + Q\mean{\Delta_0} + \frac{i \hbar}{2\tau(E)} \delta g}{\Xi \mean{\Delta_0}}.
	\label{eq:Delta/E}
\end{equation}
%

A further simplification is obtained in the limit of strongly suppressed anisotropy, $\abs{\hbar/\tau(E)} \gg \sqrt[4]{\mean{\Delta'^2}} \sqrt{\Delta_0}$.
%
%
Acting like in Sec. \ref{sub:ordinary_aniso}, we obtain a generalization of Eq. ~\eqref{eq:universal_aniso}:
\begin{equation}
	\mean{g_1} \left[ E - \mean{\Delta_0} (1 - Q) \right] - \frac{\mean{\Delta_0}}{2\mean{g_1}} + \frac{2 i \mean{\Delta'^2} \tau(E)}{\hbar} \mean{g_1}^2 = 0.
	\label{eq:universal1.5}
\end{equation}
%
This equation can describe the bound states at low $c_2$ as well as the universal smoothing with enhanced $1/\tdep$, as we will see below.

\subsection{Small concentration pairing-potential impurities}
\label{sub:small_aniso}

In contrast to the isotropic case, the pairing-potential impurities with a local gap reduction ($\Delta_1 < 0$) do not necessarily provide bound states, even in the limit of small anisotropy. 
In this Section, we derive the condition of the emergence of the bound states and evaluate the width of the impurity band in the limit of small concentrations $c_2$. 
 
The energy of the possible bound state is determined by the pole of $\check{T}_E^{(2)}$ in the limit of vanishing $c_2$,
\begin{equation}
	1 -2i f_{20} \mean{g_1(E)} = 0.
	\label{eq:bound_aniso}
\end{equation}
Let us concentrate on the limit of strongly suppressed anisotropy, described by Eq.~\eqref{eq:universal_aniso}. To satisfy Eq.~\eqref{eq:bound_aniso}, $\mean{g_1}$ must be purely imaginary. This requires that the density of states is zero at this energy, i.e., $E$ below the gap edge. We notice that in the universal limit 
the gap edge $E_{\mathrm{cr}}$ is shifted with respect to $\mean{\Delta_0}$ by a small energy\cite{Clem66PR,Skvortsov+2013JETP_subgap}
\begin{equation}
	{\cal E}_{\mathrm{cr}} \equiv \mean{\Delta_0} - E_{\mathrm{cr}} = \frac{3}{2} \mean{\Delta_0} \left( \frac{\hbar}{\mean{\Delta_0} \tdep} \right)^{2/3}.
	\label{eq:E_cr}
\end{equation}
%
At $E = E_{\mathrm{cr}}$ the averaged Green function equals
\begin{equation}
	\mean{g_1(E_{\mathrm{cr}})} = -i \left( \frac{\tdep \mean{\Delta_0}}{\hbar} \right)^{1/3},
	\label{eq:g1(Ecr)}
\end{equation}
reaching its maximal negative purely imaginary value. Hence, the existence of bound states requires
\begin{equation}
	1 -2i f_{20} \mean{g_1(E_{\mathrm{cr}})} < 0,
	\label{eq:bound_condition}
\end{equation}
or
\begin{equation}
	\frac{\hbar}{\mean{\Delta_0} \tdep} < \left(\frac{2{\cal E}_0}{\mean{\Delta_0}}\right)^{3/2},
	\label{eq:bound_existence2}
\end{equation}
where ${\cal E}_0 = 2f_{20}^2 \mean{\Delta_0}$ would give the bound state energy for an isotropic gap. This implies that even a small gap anisotropy can prevent the formation of a bound state at sufficiently small ${\cal E}_0$.

\begin{figure}[ht]
	\centering
		\includegraphics[width=\linewidth]{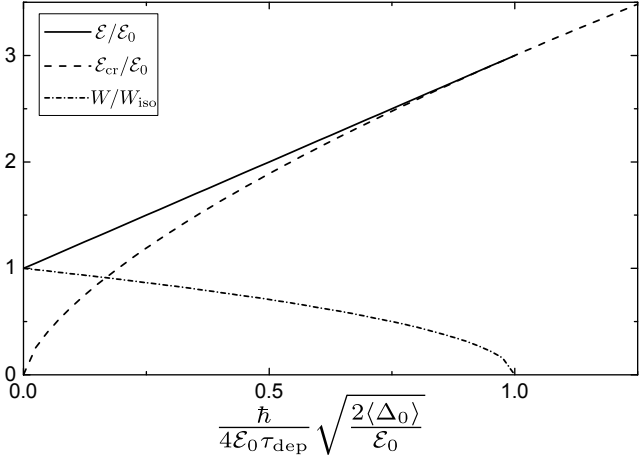}
	\caption{The bound state energy ${\cal E}$, the gap edge ${\cal E}_{\mathrm{cr}}$ and the impurity band width $W$ vs. the depairing rate $\tdep^{-1}$ in the limit of strongly suppressed anisotropy ($\tilde{P} \gg 1$).} \label{fig:band_aniso}
\end{figure}

If the bound state exists, its energy ${\cal E}$ (counted from $\mean{\Delta_0}$) is obtained by substituting $\mean{g_1(E)} = -i/(2f_{20})$ from Eq.~\eqref{eq:bound_aniso} to Eq.~\eqref{eq:universal_aniso}, and yields
%
\begin{equation}
	{\cal E} 
	= {\cal E}_0 + \frac{\hbar}{\tdep}\sqrt{\frac{\mean{\Delta_0}}{2{\cal E}_0}}.
	\label{eq:E0_aniso}
\end{equation}
The width of the impurity band $W$ in the limit of small $c_2$ is obtained from the expansion of Eq.~\eqref{eq:universal1.5} at energies close to ${\cal E}$, see Appendix \ref{app:Band}:
\begin{equation}
	W \approx 
	W_{\mathrm{iso}} \left( 1 - \frac{\hbar}{2\tdep {\cal E}_0} \sqrt{\frac{\mean{\Delta_0}}{2{\cal E}_0}} \right)^{1/2},
	\label{eq:bandwidth_aniso}
\end{equation}
%
%
%
%
where $W_{\mathrm{iso}}$ is the width of the impurity band for an isotropic gap [Eq.~\eqref{eq:W_iso}]. Equation \eqref{eq:bandwidth_aniso} is valid as long as the width of the impurity band is much smaller than the distance from an isolated bound state to the edge of the continuum, i.~e., ${\cal E} - {\cal E}_{\mathrm{cr}}$. The dependencies of the bound state energy [Eq.~\eqref{eq:E0_aniso}], the gap edge [Eq.\eqref{eq:E_cr}] and the impurity band width [Eq.~\eqref{eq:bandwidth_aniso}] on the depairing rate $\tdep^{-1}$ are shown in Fig.~\ref{fig:band_aniso}.

For comparison, we give here also the energy of the bound state and the width of the impurity band in the opposite limiting case of a vanishing concentration of ordinary scatterers. In this case, the answers are not universal, depending on the concrete shape of the anisotropic part of the pairing potential $\Delta'(\vec{n})$. We consider a common model\cite{Clem66PR}, where the values of $\Delta'$ are uniformly distributed in an interval $[-\Delta_a , \Delta_a]$. Explicit equations for $\mean{g_1}$ in this case are given in Sec.~\ref{sub:gap_numerical}. One can show that within the approximation of weak potential scattering ($\abs{f_{10}}^2 \ll 1$) a bound state exists, provided
\begin{equation}
	{\cal E}_0> \frac{\Delta_a}{2}.
	\label{eq:localize_uniform}
\end{equation}
The energy of the bound state is given by
\begin{equation}
	{\cal E} = {\cal E}_0 + \frac{\Delta_a^2}{4{\cal E}_0}.
	\label{eq:bound_uniform}
\end{equation}
The width of the impurity band at small impurity concentrations is
\begin{equation}
	W = W_{\mathrm{iso}} \left( 1 -\frac{\Delta_a^2}{4{\cal E}_0^2} \right)^{1/2}.
	\label{eq:band_uniform}
\end{equation}

\subsection{Universal behavior in the presence of pairing-potential impurities}
\label{sub:double_universal}

\begin{figure}[htb]
	\centering
		\includegraphics[width = \linewidth]{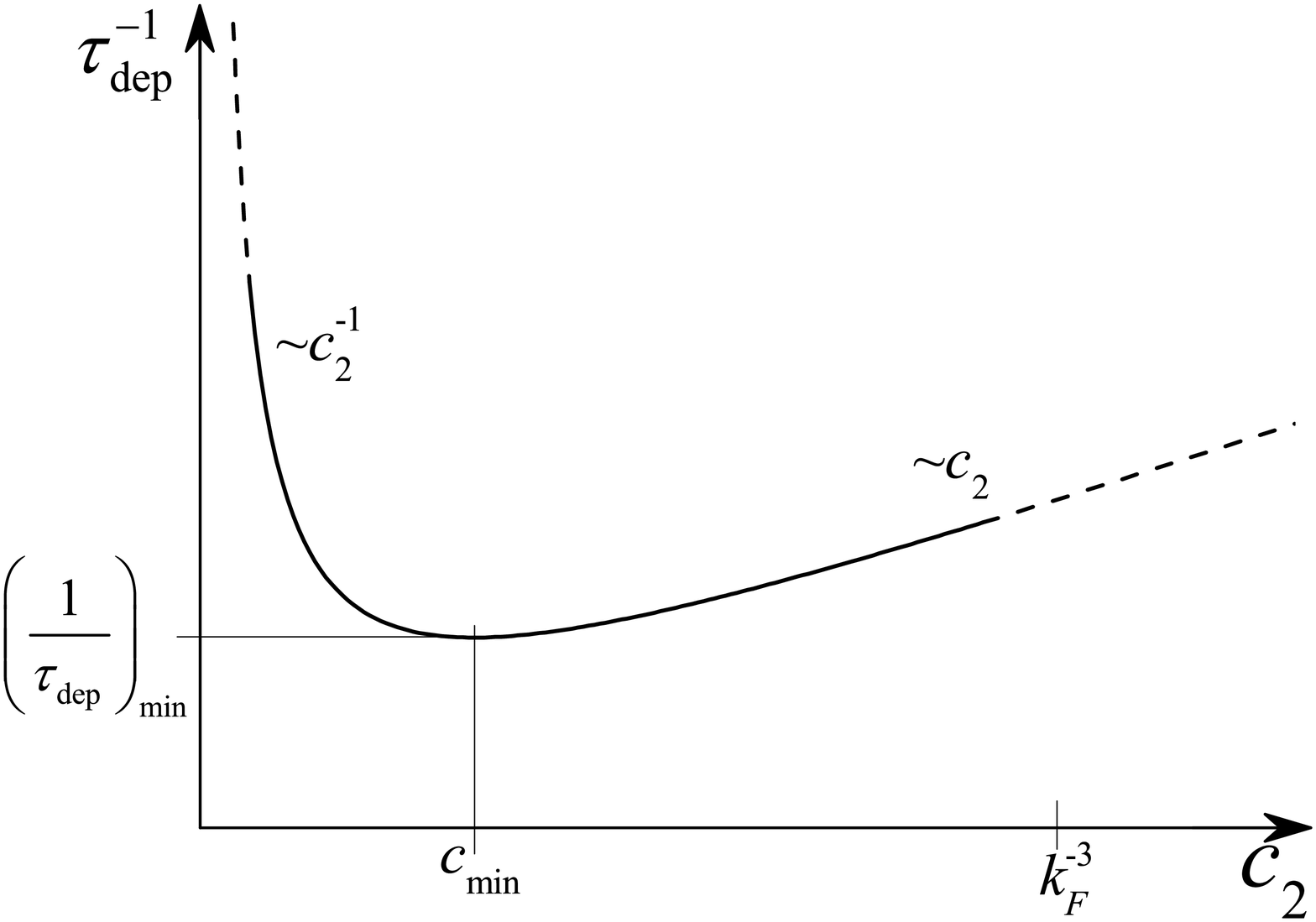}
	\caption{Schematic dependence of the depairing rate $\tdep^{-1}$ on on the concentration of pairing potential impurities in the universal limit [Eq.~\eqref{eq:double_uni_td} with $\tau_1^{-1} = 0$]. The dashed lines indicate the regions where Eq.~\eqref{eq:double_uni_td} is not applicable.}\label{fig:Depairing}
\end{figure}

As we have seen, at large impurity concentrations the shape of the smoothing of the gap edge eventually approaches the universal limit. We have discussed two situations for this to occur: the disorder in the pairing potential for an isotropic gap, and the suppression of the anisotropy of the pairing potential by potential scattering. 
Here, we consider a more general situation, where both an anisotropic gap and pairing-potential impurities are present.
The universal regime then requires 
\begin{equation}
	\abs{f_{20} \mean{g_1}} \ll 1, \quad \hbar \left( \tau_1^{-1} + \tau_2^{-1} \right) \gg \sqrt[4]{\mean{\Delta'^2}} \sqrt{\mean{\Delta_0}}.
	\label{eq:unicond1}
\end{equation}
In this limit, Eq.~\eqref{eq:universal1.5} takes the form
\begin{eqnarray}
	& \mean{g_1} \left(E - \mean{\Delta_0} + \frac{c_2 f_{20}}{\pi \nu_0} \right) - \frac{\mean{\Delta_0}}{2\mean{g_1}} & \nonumber \\
	& + 2 i \mean{g_1}^2 \left( \frac{\mean{\Delta'^2}}{\hbar (\tau_1^{-1} + \tau_2^{-1})} + \frac{c_2 f_{20}^2}{\pi \nu_0} \right)= 0. &
	\label{eq:double_uni}
\end{eqnarray}
This reproduces the universal limit with 
\begin{equation}
	\frac{1}{\tdep} =  \frac{2 \mean{\Delta'^2}}{\hbar^2 (\tau_1^{-1} + \tau_2^{-1})} + \frac{2 c_2 f_{20}^2}{\hbar \pi \nu_0},
	\label{eq:double_uni_td}
\end{equation}
and an extra shift of the gap edge $-c_2 f_{20}/(\pi \nu_0)$. Interestingly, the depairing rate exhibits a nonmonotonic dependence on the concentration $c_2$ (see Fig.~\ref{fig:Depairing}). In particular, if ordinary scatterers are absent ($P_1 = 0$), the depairing rate has a minimum at
\begin{equation}
	c_2 = c_{\mathrm{min}} \equiv \frac{\pi \nu_0 \sqrt{\mean{\Delta'^2}}}{\sqrt{2} \abs{f_{20}} \abs{f_{10}^{(2)}}}.
	\label{eq:c_min}
\end{equation}
At this concentration
\begin{equation}
	\frac{1}{\tdep} = \left( \frac{1}{\tdep} \right)_{\mathrm{min}} \equiv \frac{2 \sqrt{2 \mean{\Delta'^2}} \abs{f_{20}}}{\hbar \abs{f_{10}^{(2)}}}.
	\label{eq:min_rate}
\end{equation}
%

A major consequence of the nonmonotonicity of the depairing rate is the nonmonotonic dependence of the gap-edge smearing on the concentration of pairing-potential impurities. To ensure that this feature is present, it is sufficient to provide that the universality conditions \eqref{eq:unicond1} are satisfied at concentrations close to $c_{\mathrm{min}}$. This is the case when
\begin{equation}
  {\cal E}_0 \ll \mean{\Delta_0} \sqrt{\mean{\Delta'^2}} \abs{f_{10}^{(2)}}^2.
	\label{eq:double_uni_min}
\end{equation}
In addition, the concentration $c_{\mathrm{min}}$ should be realistic: at least, $\cimp \ll L^{-3}$. When $L \sim \lambda_F$, this yields the condition
\begin{equation}
	\frac{\mean{\Delta_0} \sqrt{\mean{\Delta'^2}}}{\mu \abs{f_2} \abs{f_{10}^{(2)}}} \ll 1.
	\label{eq:cmin_realistic}
\end{equation}

\subsection{Numerical calculations of the density of states}
\label{sub:gap_numerical}

In this section, we report some numerical calculations to examplify the typical behavior of the density of states in the presence of pairing-potential impurities and anisotropy.

Except for the universal limit, the results will depend on the details of the anisotropic part of the gap $\Delta'(\vec{n})$. To be specific and keep it simple, we concentrate on the model\cite{Clem66PR} with the values of $\Delta'$ uniformly distributed in an interval $[-\Delta_a .. \Delta_a]$.

To derive an equation for the Green function $\mean{g_1}$ in the presence of ordinary potential impurities we use Eqs.~\eqref{eq:g1_edge} and \eqref{eq:difference_edge}, where the averaging over the directions of momentum can be performed analytically. After this we eliminate $\delta g$ from the two equations to obtain the fifth-order polynomial equation
\begin{equation}
	\frac{3y^4 -2py^5}{3(1-py)^2} + 2 E_a y^2 = -1.
	\label{eq:y}
\end{equation}
Here, we made use the dimensionless variables
\begin{eqnarray}
	& y = -i \mean{g_1} \sqrt{\frac{\Delta_a}{\Delta_0}}, \qquad E_a = \frac{E - \mean{\Delta_0}}{\Delta_a},& \label{eq:y_definition} \\ 
	& p = \frac{\hbar}{2\sqrt{\Delta_a \mean{\Delta_0}} \tau_1}, &
	\label{eq:p_definition}
\end{eqnarray}
which differ from $\tilde{G}_1$, $\tilde{\Omega}$ and $\tilde{P}$, introduced in Sec. \ref{sub:ordinary_aniso}, by numeric factors.
An equation similar to Eq.~\eqref{eq:y} has already been studied by Clem\cite{Clem66PR}. To adjust the equation for the case when both pairing-potential and ordinary impurities are present, we implement Eq.~\eqref{eq:Delta/E} to show that the adjustment amounts to the following substitutions:
\begin{equation}
	p \to p_1 + \frac{p_2}{1+\alpha y}, \quad E_a \to E_a + \frac{p_2\alpha}{2\abs{f_{10}^{(2)}}^2 (1 + \alpha y)},
	\label{eq:y_substitutions}
\end{equation}
%
%
where 
\begin{equation}
	\alpha = \sqrt{\frac{2{\cal E}_0}{\Delta_a}}, \qquad p_{1,2} = \frac{\hbar}{2\sqrt{\Delta_a \mean{\Delta_0}} \tau_{1,2}}.
	\label{eq:alpha}
\end{equation}
With these substitutions Eq.~\eqref{eq:y} becomes an eigth-order polynomial equation with respect to $y$. In the limit of strongly suppressed anisotropy, this is simplified to a fifth-order polynomial equation [Eq.~\eqref{eq:universal1.5}]. We solved these equations numerically.


Let us first consider relatively large values of the parameter $\alpha$, so that the energy scale ${\cal E}_0$ is of the order of or larger than the characteristic broadening of the gap edge. The characteristic width $\Delta E_a$ of the gap edge, measured in the units $\Delta_a$, equals unity in the pure case and is of the order of $p_1^{-2/3}$ when the anisotropy is suppressed by potential scatterers: $p_1 \gg 1$. We consider the range of parameters where $\alpha^2 \gtrsim \Delta E_a$. At $\alpha^2 \gg \Delta E_a$ the broadening of the gap edge is not relevant for the formation and merging of the impurity band with the continuum. The situation is qualitatively the same as in the isotropic case, see Sec.~\ref{sec:isotropic}. It is therefore interesting to concentrate on the range of parameters $\alpha^2 \sim \Delta E_a$. 

\begin{figure}[htb]
	\centering
		\includegraphics[width = 0.49\linewidth]{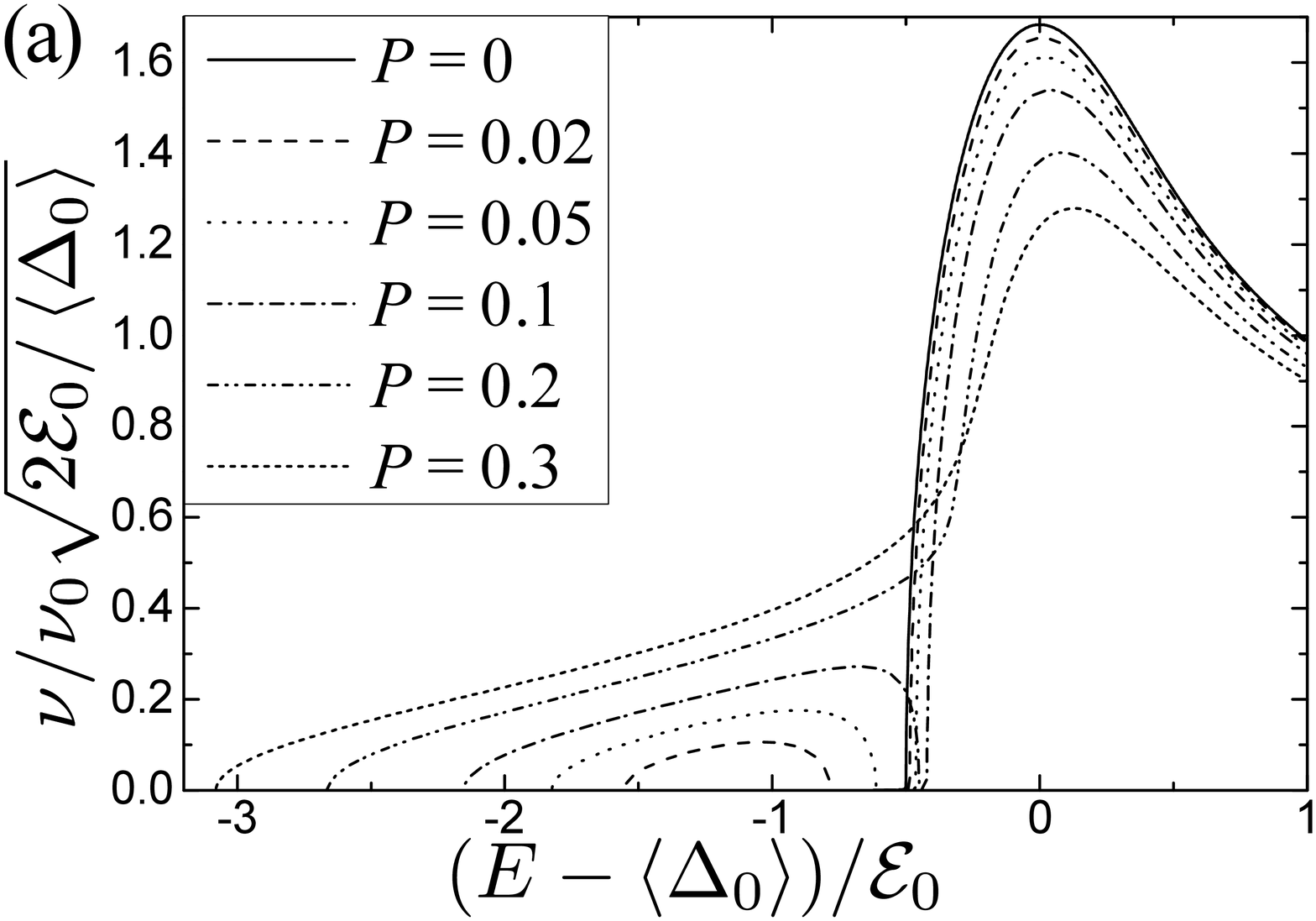}
		\includegraphics[width = 0.49\linewidth]{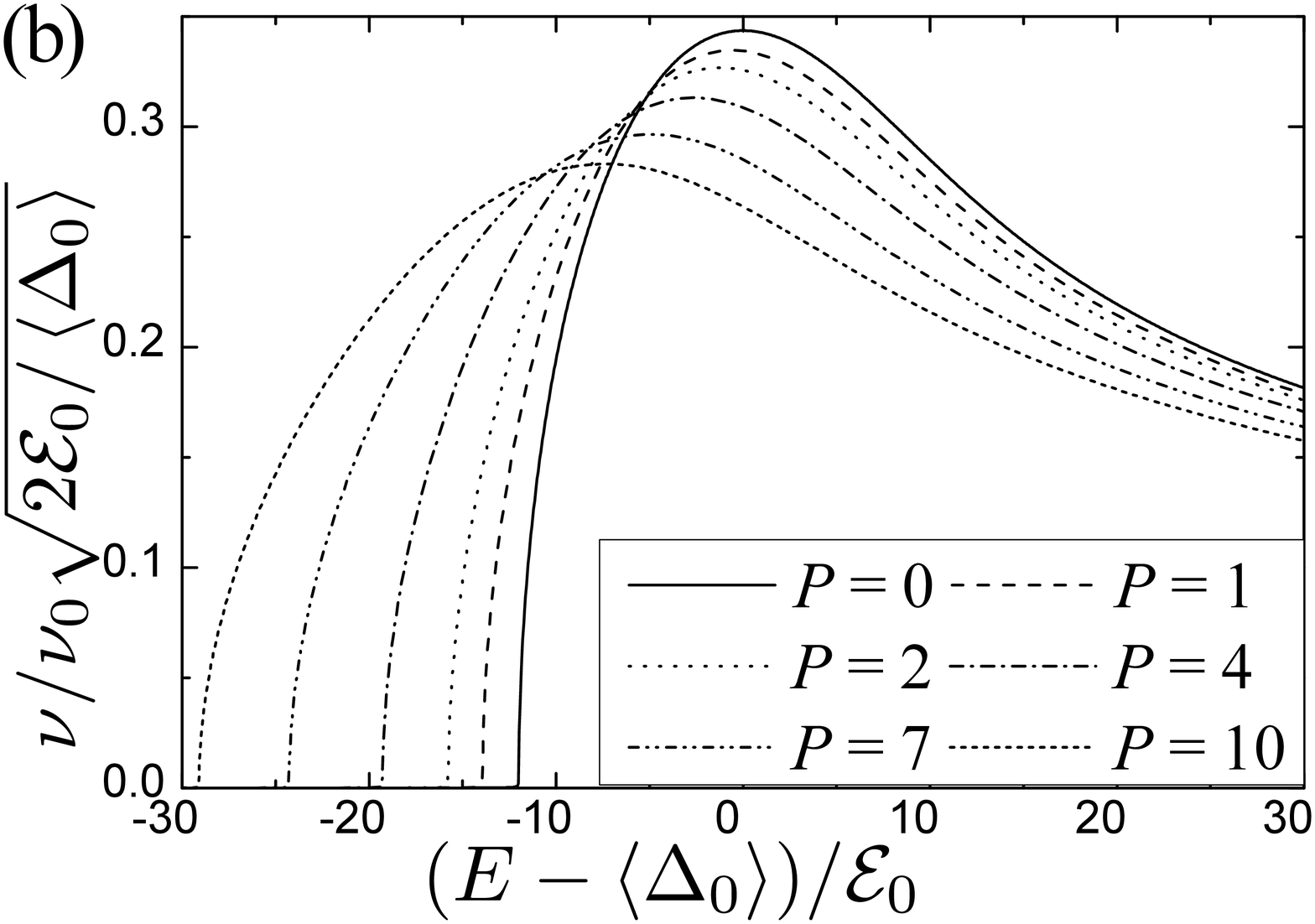}
		\includegraphics[width = 0.49\linewidth]{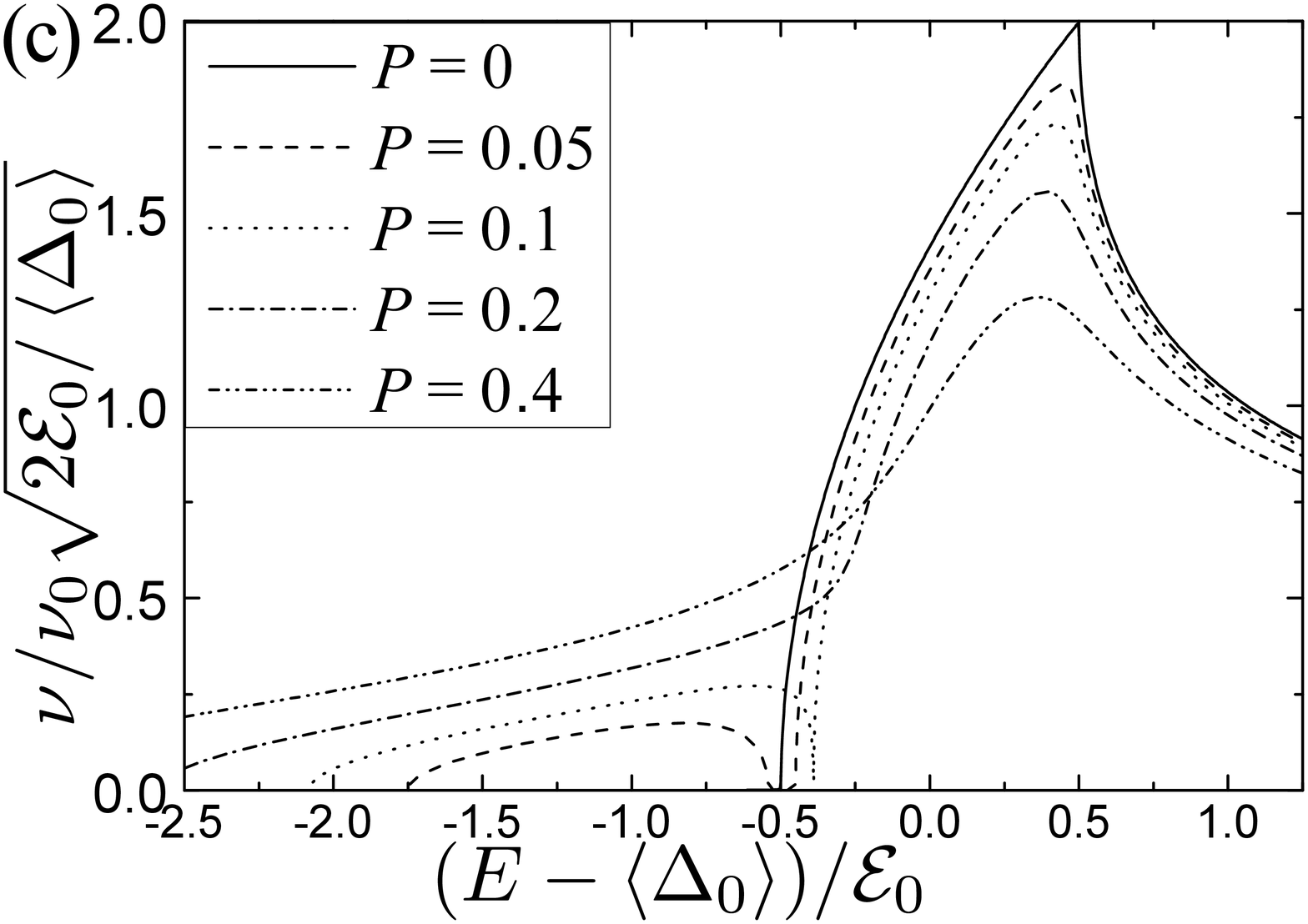}
		\includegraphics[width = 0.49\linewidth]{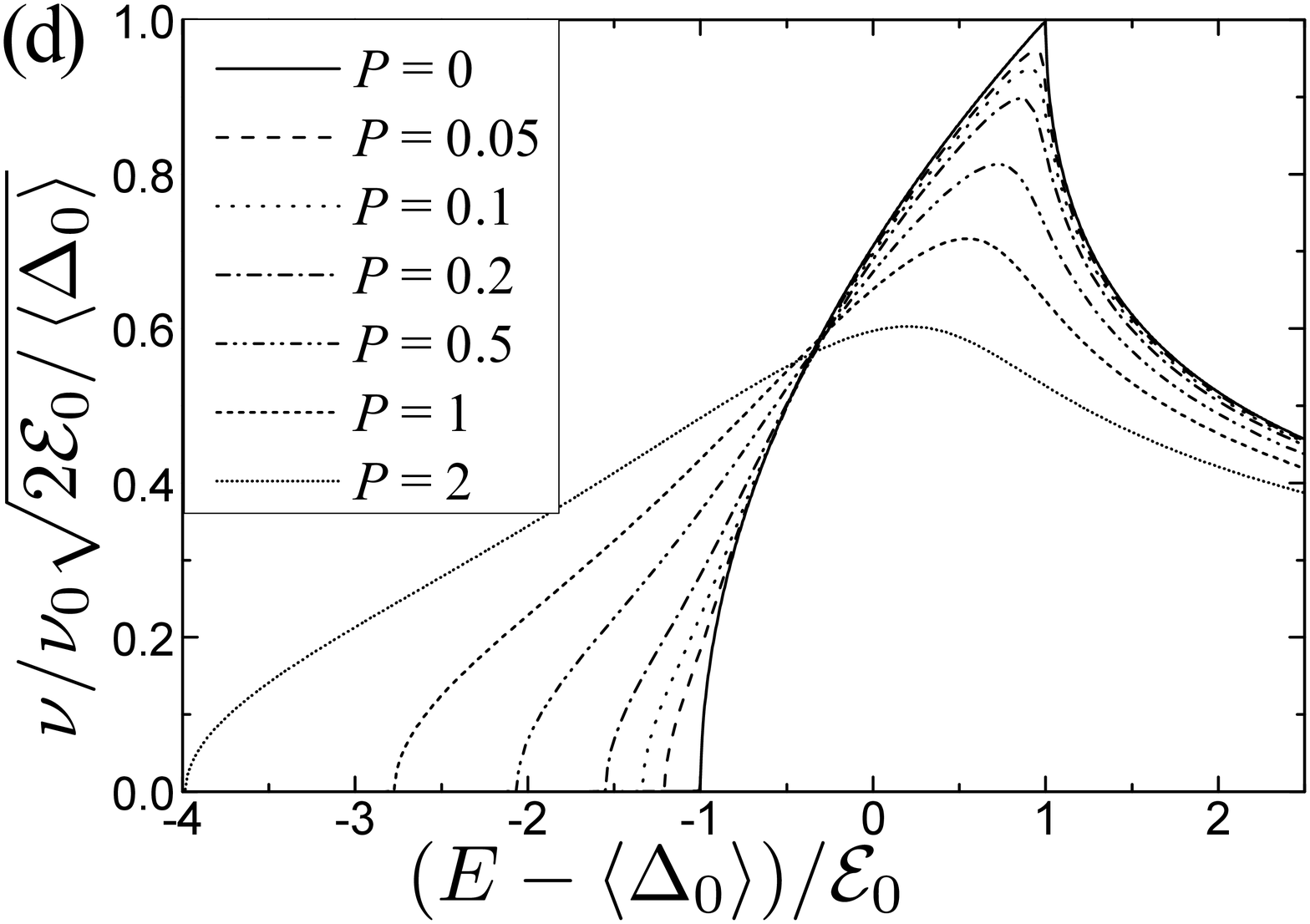}
	\caption{Density of states vs. energy in an anisotropic superconductor in the limit of strongly suppressed anisotropy (a,b) [Eq.~\eqref{eq:universal1.5}] and in the absence of ordinary potential scatterers (c,d) [Eqs.~\eqref{eq:y} and \eqref{eq:y_substitutions}]. (a) ${\cal E}_0 = 2{\cal E}_{\mathrm{cr}}$, $f_{10}^{(2)} = 0$; (b) ${\cal E}_0 = {\cal E}_{\mathrm{cr}}/8$, $f_{10}^{(2)} = 0$; (c) ${\cal E}_0 = 2\Delta_a$, $\abs{f_{10}^{(2)}}^2 = 0.1$; (d) ${\cal E}_0 = 0.5 \Delta_a$, $\abs{f_{10}^{(2)}}^2 = 0.1$.} \label{fig:Anisodensity0.1}
\end{figure}

We start with the case of strongly suppressed anisotropy, when Eq.~\eqref{eq:universal1.5} is applicable. We choose 
\begin{equation}
  \frac{\hbar}{\tdep} = \frac{{\cal E}_0}{3} \sqrt{\frac{{\cal E}_0}{3\mean{\Delta_0}}}, \quad \mbox{or} \quad \alpha^2 = 2.88 p_1^{-2/3}. 
	\label{eq:F=sqrt(6)}
\end{equation}
%
%
Then, the smoothing shifts the gap edge to $-{\cal E}_0/2$, and Eq.~\eqref{eq:E0_aniso} predicts a bound state at the energy $\epsilon \approx 1.14 {\cal E}_0$. Fig \ref{fig:Anisodensity0.1}a illustrates the behavior of the density of states upon increasing concentration of pairing potential impurities. We observe the qualitative similarity with the isotropic case -- see Fig.~\ref{fig:Density_iso}a. In particular, there is the formation of the impurity band that merges with the continuum with growing $c_2$. Moreover, the characteristic scale for the concentration $c_2$ is the same as in the isotropic situation and corresponds to $P \sim 1$, where the parameter $P$ is
\begin{equation}
	P = \frac{c_2}{\sqrt{2 {\cal E}_0 \mean{\Delta_0}} \pi \nu_0}.
	\label{eq:P_aniso}
\end{equation}
However, while in the isotropic case the smoothing of the peak was due to pairing-potential impurities only, now the pairing-potential impurities provide an extra contribution to the existing smoothing. 


The situation changes if the scales $\alpha^2$ and $\Delta E_a$ remain comparable, but the bound state is absent. For Fig.~\ref{fig:Anisodensity0.1}b we choose
\begin{equation}
	\frac{\hbar}{\tdep} = 16{\cal E}_0 \sqrt{\frac{2{\cal E}_0}{\mean{\Delta_0}}}, \quad \mbox{or} \quad \alpha^2 = 0.12 p_1^{-2/3}.
	\label{eq:F=0.5}
\end{equation}
Here, we see no trace of the impurity band. The pairing-potential impurities widen the peak and shift down the gap edge.

Apparently, the effect of potential scattering by pairing-potential impurities on the density of states is negligible, as long as 
$\alpha^2 \sim \Delta E_a$, and if the anisotropy is already suppressed by ordinary potential scatterers: $\tau_1^{-1} \gg \sqrt[4]{\mean{\Delta'^2}}\sqrt{\mean{\Delta_0}}/\hbar$. Taking $\abs{f_{10}^{(2)}} = 0$ and $\abs{f_{10}^{(2)}} = 1$ (which is the maximal permitted value) produces profiles of the density of states that are almost visually indistinguishable. To get a qualitative understand of this, let us consider, e.~g., the depairing rate in the universal limit  -- see Eq. \eqref{eq:double_uni_td}. Here, potential scatteing by pairing-potential impurities may become important only at relatively large concentrations $c_2$, when $\tau_2^{-1}$ is of the order of $\tau_1^{-1}$. However, at such values of $c_2$ the second term in the right-hand side of Eq. \eqref{eq:double_uni_td} becomes dominant, so that the depairing rate is determined by the off-diagonal scattering amlitude, $f_{20}$. Thus, the scattering rate $\tau_2^{-1}$ can be neglected in Eq. \eqref{eq:double_uni_td} at all concentrations $c_2$.


It is interesting to spot the similarities with the seemingly different situation, when potential scatterers are absent --- see Figs.~\ref{fig:Anisodensity0.1}c,d. Here, we use Eqs.~\eqref{eq:y} and \eqref{eq:y_substitutions}, putting $p_1 = 0$. The density of states at $p_2 = 0$ exhibits a typical cusp structure, typical for the model we use. However, the qualitative behavior of the density of states is analogous to the case of strongly suppressed anisotropy, if we choose ${\cal E}_0 \sim \Delta_a$ ($\alpha^2 \sim 1$). If the shift of the gap edge amounts to ${\cal E}_0/2$, the bound state is formed at $\epsilon = 1.06{\cal E}_0$ -- see Fig.~\ref{fig:Anisodensity0.1}c. Upon increasing the impurity concentration, we see again the impurity band formation, merging with the continuum, and the extra smoothing of the coherence peak at the scales $P \sim 1$ of the dimensionless concentration. For Fig.~\ref{fig:Anisodensity0.1}d we choose ${\cal E}_0 = \Delta_a/2$. This corresponds to the theshold of the bound state formation. Similar to Fig.~\ref{fig:Anisodensity0.1}b, we don't see any signs of the impurity band, rather the effect is a combination of the peak smoothing and shift.

\begin{figure}[htb]
	\centering
		\includegraphics[width = \linewidth]{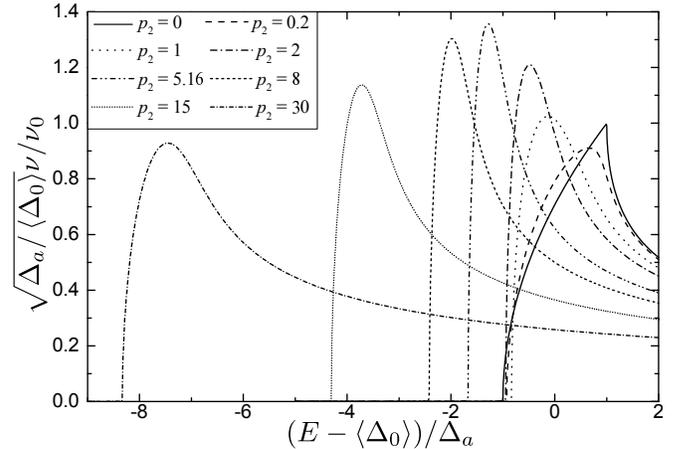}
	\caption{Density of states vs. energy in an anisotropic superconductor containing pairing-potential impurities with a small parameter ${\cal E}_0$: ${\cal E}_0 = 0.025 \abs{f_{10}}^2 \Delta_a$, $\abs{f_{10}^{(2)}}^2 = 0.1$.} \label{fig:Aniso_Width}
\end{figure}

Finally, we have modeled the density of states in another interesting limiting case, when $\alpha$ is very small: $\alpha^2 \ll \abs{f_{10}^{(2)}}^2$. To remind, this inequality guarantees the nonmonotonic behavior of the peak smoothing vs. the impurity concentration. In this case, at a small impurity concentration the pairing-potential impurities affect the anisotropic part of $\Delta_0$ mostly as potential scatterers. One can see this in Fig.~\ref{fig:Aniso_Width}, where the incerease of the dimensionless impurity concentration $p_2$ results in the narrowing of the peak. Further increase of the concentration gives the peak shift, manifesting the pairing potential distortion brought by the impurities. Starting from $p_2 = 5.16$ the peak also becomes wider upon increasing concentration, in accordance with Eq.~\eqref{eq:double_uni_td}.

\section{Conclusion} 
\label{sec:Conclusion}

In conclusion, we have considered the effect of pairing-potential impurities on the behavior of the density of states in a superconductor. We found that this behavior is strongly affected by the anisotropy of the bulk pairing potential.

We have considered first the limit of negligible anisotropy. In this case, we established an analogy between the pairing-potential impurities and weak polarized magnetic impurities. This analogy allows to extend the results obtained for one type of defects to the other type. The persistence of bound states at single impurities is typical for the isotropic case. We demonstrate that a spherically symmetric impurity with local suppression of $\Delta$ gives rise to an infinite number of subgap bound states. Upon increasing the impurity concentration, these states form an impurity band that eventually merges with the continuum, resulting in a smoothed gap edge.

Even a slightly anisotropic pairing potential forbids the formation of bound states at sufficiently small pairing-potential distortion at the impurities. We derived the criterion of existence of the bound states and have analyzed in detail the behavior of the density of states.

\appendix
\section{}
\label{app:T}

In this Appendix we derive Eqs. \eqref{eq:commute}, \eqref{eq:det} and \eqref{eq:T_main}. 

Equations \eqref{eq:main} and \eqref{eq:principal} yield
\begin{equation}
	\check{g}(E,\vec{n}) = i \int \left[ \check{S}(E,p\vec{n}) - \check{1} \xi(p) \right]^{-1} \frac{d\xi(p)}{\pi},
	\label{eq:g_through_S}
\end{equation}
where 
\begin{equation}
	\check{S}(E,\vec{p}) = \left(
	\begin{array}{cc}
	  E + i\epsilon^+ & \Delta_0(\vec{p}) \\
		-\Delta_0(\vec{p}) & -E - i\epsilon^+
	\end{array} \right)
	-\cimp \check{T}_E(\vec{p},\vec{p}).
	\label{eq:S(p)}
\end{equation}
Let us assume that we can keep $\check{S}(E,\vec{p})$ constant when integrating over $\xi$, i. e., put $\check{S}(E,p\vec{n}) \approx \check{S}(E,k_F\vec{n}) \equiv \check{S}_E(\vec{n})$. Such simplification is justified if $\check{S}(E,p\vec{n})$ does not change significantly while $\xi(p)$ is of the order of or smaller than the largest of the moduli of the eigenvalues of $\check{S}_E(\vec{n})$. Then, we have
%
%
\begin{equation}
	\check{g}(E,\vec{n}) \approx i\dashint_{-\infty}^{+\infty} \left[ \check{S}_E(\vec{n}) - \check{1} \xi \right]^{-1} \frac{d\xi}{\pi}.
	\label{eq:main_integrated}
\end{equation}
%
From this relation immediately follows Eq. \eqref{eq:commute}. By writing Eq. \eqref{eq:main_integrated} in the basis where the matrix $\check{S}_E(\vec{n})$ is diagonal it can be proven that each eigenvalue of $\check{g}$ can be either $1$ or $-1$. In the pure case
\begin{equation}
	\check{g}(E,\vec{n}) = \frac{i}{\sqrt{\abs{\Delta_0(\vec{n})}^2 - E^2}} \left(
	\begin{array}{cc}
	  -E & -\Delta_0(\vec{n}) \\
		\Delta_0(\vec{n}) & E
	\end{array} \right),
	\label{eq:pure}
\end{equation}
and the two eigenvalues are $1$ and $-1$. There is no reason for them to change discontinuously with growing impurity concentration, hence, for any value of $\cimp$ we have $\mathrm{Tr} \, \check{g} = 0$, and $\mathrm{det} \, \check{g} = 1$, which gives Eq. \eqref{eq:det}.

To simplify Eq. \eqref{eq:T}, we want to exclude the region of integration far from the Fermi surface. To do this, we employ the trick used in Ref. \onlinecite{Rusinov1969JETP}. Let us introduce an auxiliary normal state scattering matrix $\check{f}(\vec{p},\vec{p}')$, defined by Eq. \eqref{eq:f}. 
%
%
%
%
Using Eqs. \eqref{eq:T} and \eqref{eq:f}, we obtain
\begin{eqnarray}
	& \check{T}_E(\vec{p},\vec{p}') = \check{f}(\vec{p},\vec{p}') -\int \check{f}(\vec{p},\vec{k}) \check{\calG}(k) \Vimp(\vec{k} - \vec{p}') \frac{d^3 \vec{k}}{(2\pi)^3} & \nonumber \\
	& + \int \check{f}(\vec{p},\vec{k}) \check{G}_{E}(\vec{k}) \check{T}_E(\vec{k},\vec{p}') \frac{d^3 \vec{k}}{(2\pi)^3} & \nonumber \\
	& - \int \!\! \int \check{f}(\vec{p},\vec{k}) \check{\calG}(k) \Vimp(\vec{k} - \vec{k}') \check{G}_{E}(\vec{k}') \check{T}_E(\vec{k}',\vec{p}') \frac{d^3 \vec{k}}{(2\pi^3)} \frac{d^3 \vec{k}'}{(2\pi)^3} & \nonumber \\
	& \hspace{-0.6cm} = \check{f}(\vec{p},\vec{p}') \! + \! \int \check{f}(\vec{p},\vec{k}) (\check{G}_E(\vec{k}) \! - \check{\calG}(k)) \check{T}_E(\vec{k},\vec{p}') \frac{d^3 \vec{k}}{(2\pi)^3}. &
	\label{eq:Af}
\end{eqnarray}
Far from the Fermi surface the difference between $\check{G}_{E}(\vec{k})$ and $\check{\calG}(k)$ vanishes. This typically happens at $\abs{\xi(k)} \gg \max(\abs{\Delta_0(\vec{n})})$, $\abs{E}$ (strictly speaking, in this estimate the quantity $\xi(p)$, defined in Eq. \eqref{eq:xi}, should include the renormalized chemical potential due to the impurities. This renormalization can be much larger than $\Delta_0$, however, as long as it is much smaller than $\mu$, it has a negligible effect on the results an will be disregarded further). For impurities with a size much smaller than the coherence length $\xi_S$ the functions $\check{f}(\vec{p},\vec{k})$ and $\check{T}_E(\vec{k},\vec{p}')$ depend very weakly on $\abs{\vec{k}}$ as long as $\xi(k) \sim \Delta_0$ (see Appendix \ref{app:f}). This means that we can integrate in Eq. \eqref{eq:Af} over $\abs{\vec{k}}$ and obtain Eq. \eqref{eq:T_main}. 
%

\section{}
\label{app:f}

In this appendix the matrix $\check{f}(\vec{p},\vec{p}')$ will be evaluated for a spherically symmetric impurity. Considering Eqs. \eqref{eq:f} and \eqref{eq:Vimp_iso} with a real function $\Delta_1(\vec{p})$ one can see that $\check{f}$ has the form
\begin{equation}
	\check{f}(\vec{p},\vec{p}') = \left(
	\begin{array}{cc}
    f_1(\vec{p},\vec{p}') & f_2(\vec{p},\vec{p}') \\
	  -f_2^*(\vec{p},\vec{p}') & f_1^*(\vec{p},\vec{p}')
\end{array} \right). 
\label{eq:f_components1}
\end{equation}
The equations for $f_1$ and $f_2$ read
\begin{eqnarray}
	& f_1(\vec{p},\vec{p}') = U(\vec{p} - \vec{p}') + \int \left[ f_1(\vec{p},\vec{k}) G_N^{(0)}(k) U(\vec{k} - \vec{p}') \right. & \nonumber \\
	& \left. + f_2(\vec{p},\vec{k}) G_N^{(0)*}(k) \Delta_1(\vec{k} - \vec{p}') \right] \frac{d^3 \vec{k}}{(2\pi)^3}, &
	\label{eq:f_11}
\end{eqnarray}
\begin{eqnarray}
	& f_2(\vec{p},\vec{p}') = -\Delta_1(\vec{p} - \vec{p}') + \int \left[ f_2(\vec{p},\vec{k}) G_N^{(0)*}(k) U(\vec{k} - \vec{p}') \right. & \nonumber \\
	& \left. - f_1(\vec{p},\vec{k}) G_N^{(0)}(k) \Delta_1(\vec{k} - \vec{p}') \right] \frac{d^3 \vec{k}}{(2\pi)^3}, &
	\label{eq:f_12}
\end{eqnarray}
where
\[ G_N^{(0)}(k) = (-\xi(k) +i\epsilon^+)^{-1}. \]
The functions $f_1(\vec{p},\vec{p}')$ and $f_2(\vec{p},\vec{p}')$ are the ordinary and Andreev scattering amplitudes, respectively, for an electron in the normal state incident at an impurity with an electric potential $U(\vec{r})$ and pairing potential $\Delta_1(\vec{r})$. To solve Eqs. \eqref{eq:f_11} and \eqref{eq:f_12}, we assume that Andreev scattering can be taken into account within the perturbation theory. This is the case when
\begin{equation}
	\frac{k_F}{\mu} \int \abs{\Delta_1(r)} dr \ll 1,
	\label{eq:Andreev_small}
\end{equation}
if $\abs{U(\vec{r})} \lesssim \mu$. Note that for $\abs{\Delta_1} \sim \Delta_0$ Eq. \eqref{eq:Andreev_small} simply means that $L \ll \xi_S$. Then, the last term in the right-hand side of Eq. \eqref{eq:f_11}, which is second order in $\Delta_1$, can be neglected. As a result, $f_1 \approx f_N$, where $f_N$ is the vertex part of the normal-state Green function in the presence of a single impurity:
\begin{equation}
	f_N(\vec{p},\vec{p}') = U(\vec{p} - \vec{p}') + \int f_N(\vec{p},\vec{k}) G_N^{(0)}(k) U(\vec{k} - \vec{p}') \frac{d^3 \vec{k}}{(2\pi)^3}.
	\label{eq:f_N}
\end{equation}
The differential scattering cross-section $d\sigma$ on the potential $U(r)$ from $\vec{p}_1$ and $\vec{p}_2$ (both lying on the Fermi surface) is
\begin{equation}
	\frac{d\sigma}{d\Omega} (\vec{p}_1 \to \vec{p}_2) = \frac{m^2}{4\pi^2 \hbar^4} \abs{f_N(\vec{p}_2,\vec{p}_1)}^2.
	\label{eq:sigma_diff}
\end{equation}

To solve Eq. \eqref{eq:f_12} we note that the function
\begin{equation}
	\Gamma(\vec{p},\vec{p}') = \delta(\vec{p}-\vec{p}') + f_N^*(\vec{p},\vec{p}') \frac{G_N^{(0)*}(p)}{(2\pi)^3}
	\label{eq:Gamma}
\end{equation}
satisfies the equation
\begin{equation}
	\Gamma(\vec{p},\vec{p}') = \delta(\vec{p} - \vec{p}') + \int \Gamma(\vec{p},\vec{k}) G_N^{(0)*}(k) U(\vec{k} - \vec{p}') \frac{d^3 \vec{k}}{(2\pi)^3}.
	\label{eq:Gamma_Green}
\end{equation}
Hence,
\begin{eqnarray}
  & f_2 \approx \int \left[ -\Delta_1(\vec{p} - \vec{q}) \right. & \nonumber \\
	& \left. -\int f_N(\vec{p},\vec{k}) G_N^{(0)}(k) \Delta_1(\vec{k} - \vec{q})  \frac{d^3 \vec{k}}{(2\pi)^3}\right] \Gamma(\vec{q},\vec{p}') d^3 \vec{q} & \nonumber \\
	& = - \Delta_1(\vec{p} - \vec{p}') - \int f_N(\vec{p},\vec{k}) G_N^{(0)}(k) \Delta_1(\vec{k} - \vec{p}') \frac{d^3 \vec{k}}{(2\pi)^3} & \nonumber \\
	& - \int \Delta_1(\vec{p} - \vec{k}) G_N^{(0)*}(k) f_N^*(\vec{k},\vec{p}') \frac{d^3 \vec{k}}{(2\pi)^3} & \nonumber \\
	& - \! \int \!\! f_N(\vec{p},\vec{k}) G_N^{(0)}(k) \Delta_1 (\vec{k} \! - \!\vec{q}) G_N^{(0)*}\!\!(q) f_N^*(\vec{q},\vec{p}') \frac{d^3 \vec{k}}{(2\pi)^3} \frac{d^3 \vec{q}}{(2\pi)^3}. & \nonumber \\
	\label{eq:f12_solution}
\end{eqnarray}

The normal-state single-impurity Green function is given by
\begin{equation}
	G_N(\vec{p},\vec{p}') = \delta(\vec{p} - \vec{p}') G_N^{(0)}(\vec{p}) + G_N^{(0)}(\vec{p}) \frac{f_N(\vec{p},\vec{p}')}{(2\pi)^3} G_N^{(0)}(\vec{p}').
	\label{eq:G_Np}
\end{equation}
Hence,
\begin{eqnarray}
	& f_2(\vec{p},\vec{p}') = - \int G_N^{(0)-1}(p) G_N(\vec{p},\vec{k}) \Delta_1 (\vec{k} - \vec{q}) G_N^*(\vec{q},\vec{p}') & \nonumber \\
	& \times G_N^{(0)*-1}(p') d^3 \vec{k} d^3 \vec{q} = -\int e^{i\vec{p}\vec{r} - i\vec{p}'\vec{r}'} G_N^{(0)-1}(p) & \nonumber \\
	& \hspace{-0.9cm} \times G_N(\vec{r},\vec{r}_1 \!) \Delta_1 (\vec{r}_1) G_N^*(\vec{r}_1,\vec{r}') G_N^{(0)*-1}\! (p') d^3 \vec{r} d^3 \vec{r}_1 d^3 \vec{r}'\!\!. &
	\label{eq:f_12G_N}
\end{eqnarray}
Here we used that $G_N(-\vec{r}_1,-\vec{r}') = G_N(\vec{r}_1,\vec{r}')$ due to inversion symmetry. The function $G_N(\vec{r},\vec{r}')$ is defined by the equation
\begin{equation}
	\left( -\frac{\hbar^2}{2m} \frac{\partial^2}{\partial \vec{r}^2} - \mu + U(r) - i\epsilon^+ \right) G_N(\vec{r},\vec{r}') = -\delta(\vec{r} - \vec{r}').
	\label{eq:G_N}
\end{equation}
Now we expand $G_N$ and the $\delta$-function in Legendre polynomials $P_l$:
\begin{equation}
	G_N(\vec{r},\vec{r}') = \sum_l P_l \left(\frac{\vec{r}}{r} \cdot \frac{\vec{r}'}{r'} \right) \frac{G_l(r,r')}{r},
	\label{eq:G_N_expand}
\end{equation}
\begin{equation}
	\delta(\vec{r} - \vec{r}') = \frac{\delta(r - r')}{r^2} \sum_{l=0}^{\infty} \frac{2l+1}{4\pi} P_l \left(\frac{\vec{r}}{r} \cdot \frac{\vec{r}'}{r'} \right).
	\label{eq:delta_expand}
\end{equation}
%
Then
\begin{eqnarray}
	& \left[ -i\epsilon^+ + \frac{\hbar^2 l(l+1)}{2mr^2} - \frac{\hbar^2}{2m} \frac{\partial^2}{\partial r^2} - \mu + U(r) \right] G_l(r,r') & \nonumber \\
	& = -\frac{2l+1}{4\pi r'} \delta(r-r'). &
	\label{eq:Gl}
\end{eqnarray}
Let us denote as $u_{l0}$ the solution of the homogeneous equation (without the right-hand side) having the asymptotics
\begin{equation}
	u_{l0} = e^{ik_F r- \epsilon^+ k_F r/2\mu}
	\label{eq:u_asympt}
\end{equation}
at $r \to \infty$. It should be noted that at $r > r'$ $G_l(r,r')$ is proportional to $u_{l0}$, since the second linear independent solution of the homogeneous equation diverges at $r \to \infty$. Then, using standard methods for solving linear inhomogeneous equations (e. g., the method of variation of parameters) one can express $G_l$ in terms of $u_{l0}$ and $u_{l0}^*$:
\begin{equation}
	G_l(r,r') = \left\{
	\begin{array}{l}
	  s(r') u_{l0}(r) - \frac{im}{k_F \hbar^2} u_{l0}(r') \frac{2l+1}{4\pi r'} u_{l0}^*(r), \hspace{0.2cm} r<r' \\
		\left[ s(r') - \frac{im}{k_F \hbar^2} u_{l0}^* (r') \frac{2l+1}{4\pi r'} \right] u_{l0}(r), \qquad r>r'
	\end{array} \right.
	\label{eq:Gl'_solution}
\end{equation}
Here $s(r')$ is some function that will be determined from the boundary condition at $r=0$. When deriving Eq. \eqref{eq:Gl'_solution} it has been used that the Wronskian
\begin{equation}
	W = \left|
	\begin{array}{cc}
	  u_{l0}(r) & u_{l0}^*(r) \\
		u_{l0}'(r) & {u_{l0}^*}'(r)
	\end{array} \right|
	\label{eq:W}
\end{equation}
is almost constant: indeed, it can be demonstrated using Eq. \eqref{eq:Gl} (without the right-hand side) that $dW/dr \sim \epsilon^+$, and hence $W(r) \approx -2ik_F$ for not too large $r$.

To determine $s(r')$ we use the boundary condition $G_l(0,r') = 0$:
\begin{equation}
	s(r') = \frac{im}{\hbar^2 k_F} u_{l0}(r') \frac{2l+1}{4\pi r'} c_l,
	\label{eq:s}
\end{equation}
where
\[ c_l = \lim_{r \to 0} \frac{u_{l0}^*(r)}{u_{l0}(r)}. \]
Then
\begin{equation}
	G_l(r,r') = \frac{im}{\hbar^2 k_F} \frac{2l+1}{4\pi r'} \left\{
	\begin{array}{l}
	  u_l(r') u_{l0}(r), \qquad r>r' \\
		u_l(r) u_{l0}(r'), \qquad r'>r
	\end{array} \right. ,
	\label{eq:Gl_solution}
\end{equation}
where
\begin{equation}
	u_l(r) = c_l u_{l0}(r) - u_{l0}^*(r).
	\label{eq:u_l}
\end{equation}
Now we return to Eq. \eqref{eq:f_12G_N}. For further transformations we will use the addition theorem
\begin{equation}
	P_l\left(\frac{\vec{r}}{r} \cdot \frac{\vec{r}'}{r'} \right) = \frac{4\pi}{2l+1} \sum_{m=-l}^{m=l} Y_{lm}\left(\frac{\vec{r}}{r} \right) Y_{lm}^*\left(\frac{\vec{r}'}{r'} \right),
	\label{eq:addition}
\end{equation}
and the expansion [see Ref. \onlinecite{Abramowitz}]
\begin{equation}
	e^{i\vec{p}\vec{r}} = \sum_{l=0}^{\infty} (2l+1) i^l j_l(pr) P_l \left(\frac{\vec{r}}{r} \cdot \vec{n} \right).
	\label{eq:exp_expand}
\end{equation}
Here $Y_{lm}$ are the spherical harmonics and $j_l$ are the spherical Bessel functions, which are related to ordinary Bessel functions $\mathrm{J}_{\nu}$ via
\begin{equation}
	j_l(x) = \sqrt{\frac{\pi}{2x}} \mathrm{J}_{l + \frac{1}{2}} (x).
	\label{eq:Bessel_Bessel}
\end{equation}
 
If one expands the Legendre polynomials in Eq. \eqref{eq:G_N_expand} in spherical harmonics, one can perform integration over the directions of $\vec{r}$, $\vec{r}'$ and $\vec{r}_1$ in Eq. \eqref{eq:f_12G_N}:
\begin{eqnarray}
	& f_2(\vec{p},\vec{p}') = -G_N^{(0)-1}(p) G_N^{(0)*-1}(p') \sum\limits_{l=0}^{\infty} P_l(\vec{n} \cdot \vec{n}') \frac{(4\pi)^3}{2l+1} & \nonumber \\
  & \times  \int j_l(pr) j_l(p'r') G_l(r,r_1) \Delta_1(r_1) G_l^*(r_1,r') r r_1 r'^2 dr dr_1 dr'. & \nonumber \\ 
	\label{eq:f12_noAngles}
\end{eqnarray}
Further simplifications are possible if we take $\vec{p}$ and $\vec{p'}$ sufficiently close to the Fermi surface: $\abs{p-k_F}L \ll 1$ and $\abs{p'-k_F}L \ll 1$. At such parameters we may neglect the contribution to the integral in \eqref{eq:f12_noAngles} from  the region where either $r<L$ or $r'<L$ as compared to the contribution from the region where both $r>L$ and $r'>L$ (this statement is proved by the estimates given below, in particular, Eq. \eqref{eq:jlu_integral}). Then, since at $r_1>L$ $\Delta_1(r_1)$ is negligible, we may put $r_1<L<r,r'$ in Eq. \eqref{eq:f12_noAngles}: 
\begin{eqnarray}
	& f_2(\vec{p},\vec{p}') = -G_N^{(0)-1}(p) G_N^{(0)*-1}(p') \frac{4\pi m^2}{\hbar^4 k_F^2}& \nonumber \\
  & \times \sum\limits_{l=0}^{\infty} (2l+1) P_l(\vec{n} \cdot \vec{n}') \int\limits_L^{\infty} j_l(pr) u_{l0}(r) r dr & \nonumber \\
	& \times  \int\limits_L^{\infty} j_l(p'r') u_{l0}^*(r') r' dr' \int\limits_0^L \abs{u_l(r_1)}^2 \Delta(r_1) dr_1. & 
	\label{eq:f12_large_r}
\end{eqnarray}
The asymtotic behavior of $j_l(x)$ at $x \to \infty$ is
\begin{equation}
	j_l (x) \approx \frac{1}{x} \cos \left[ x - (l+1) \frac{\pi}{2} \right].
	\label{eq:jl_asympt}
\end{equation}
Then, using Eq. \eqref{eq:u_asympt}, we obtain
\begin{eqnarray}
  & \int\limits_L^{\infty} j_l(pr) u_{l0}(r) r dr \approx \int\limits_L^{L'} j_l(pr) u_{l0}(r) r dr & \nonumber \\
	& + \frac{i^{l+1} e^{i(k_F -p)L'}}{2p} \left[ \frac{k_F \epsilon^+}{2\mu} - i(k_F - p) \right]^{-1}, &
	\label{eq:jlu_integral}
\end{eqnarray}
where $L'$ is the characteristic distance at which the asymptotics \eqref{eq:jl_asympt} can be used (if this distance is smaller than $L$, we take $L'=L$). To ensure that $f_2(\vec{p},\vec{p}')$ weakly depends on $\abs{\vec{p}}$ and $\abs{\vec{p}'}$ when $\xi(p)\lesssim \Delta_0$ and $\xi(p') \lesssim \Delta_0$, we need to demand that the argument of the exponent in the second line of Eq. \eqref{eq:jlu_integral} is small, i. e., $L' \abs{k_F - p} \sim L'/\xi_S \ll 1$. According to Ref. \onlinecite{Abramowitz}, the asymptotics \eqref{eq:jl_asympt} works at $x \gg l^2$. Thus, $L' \sim l^2/k_F$, so we demand
\begin{equation}	
	l^2 \ll k_F \xi_S.
	\label{eq:l_restriction}
\end{equation}
Then, we have
\begin{equation}
  \int\limits_L^{\infty} j_l(pr) u_{l0}(r) r dr \approx \frac{i^{l+1}}{2p} \left[ \frac{k_F \epsilon}{2\mu} - i(k_F - p) \right]^{-1}.
	\label{eq:jlu_integral1}
\end{equation}
Finally, Eqs. \eqref{eq:f12_large_r} and \eqref{eq:jlu_integral1} yield
\begin{equation}
	f_2(\vec{n},\vec{n}') \approx -\frac{\pi^2 \nu_0}{k_F^2} \sum_{l=0}^{\infty} (2l+1) P_l(\vec{n} \cdot \vec{n}') \!\! \int_0^{\infty} \!\!\!\! \abs{u_l(r)}^2 \Delta_1(r) dr.
	\label{eq:f12_final}
\end{equation}
Note that the condition \eqref{eq:Andreev_small} provides the smallness of the expansion coefficients: $\abs{f_{2l}} \ll 1$.

In the end of the appendix we will derive some more useful relation for $\check{f}$. It follows from Eq. \eqref{eq:f} that
\begin{eqnarray*}
  & \Im[\check{f}(\vec{p},\vec{p}')] = \int \Im [\check{f}(\vec{p},\vec{k})] \hat{\calG}(k) \Vimp(\vec{k}-\vec{p}') \frac{d^3 \vec{k}}{(2\pi)^3} & \\
	& + \frac{i}{2} \int \check{f}^*(\vec{p},\vec{k}) [ \hat{\calG}^*(k) - \hat{\calG}(k)] \Vimp(\vec{k} - \vec{p}') \frac{d^3 \vec{k}}{(2\pi)^3}& \\
	& = \int \Im [\check{f}(\vec{p},\vec{k})] \hat{\calG}(k) \Vimp(\vec{k}-\vec{p}') \frac{d^3 \vec{k}}{(2\pi)^3} & \\
	& + \pi \nu_0 \int \check{f}^*(\vec{p},k_F \vec{n}) \left(
	\begin{array}{cc}
	  -1 & 0 \\
		0 & 1
	\end{array} \right)
	\Vimp(k_F \vec{n} - \vec{p}') \frac{d\vec{n}}{4\pi}. &
\end{eqnarray*}
Then, using Eq. \eqref{eq:f} it can be proved that
\begin{equation}
	\Im[\check{f}(\vec{n},\vec{n}')] = \int \check{f}^*(\vec{n},\vec{n}'') \left(
	\begin{array}{cc}
	  -1 & 0 \\
		0 & 1
	\end{array} \right)
	\check{f} (\vec{n}'',\vec{n}') \frac{d\vec{n}''}{4\pi}.
	\label{eq:Im(f)}
\end{equation}
Substituting here Eq. \eqref{eq:f_expand}, after integration over $\vec{n}''$ one finds that
\begin{eqnarray*}
 & \Im(f_{1l}) = -\abs{f_{1l}^2} - f_{2l}^{*2} = -\abs{f_{1l}^2} - f_{2l}^{2}, & \\
 & \Im(f_{2l}) = -2i f_{1l}^* \Im(f_{2l}) =  2i f_{1l} \Im(f_{2l}). &
\end{eqnarray*}
Finally, we obtain the following relations for the imaginary parts of $f_{1l}$ and $f_{2l}$:
\begin{equation}
	\Im (f_{1l}) = - \mathrm{det} \check{f}_l, \qquad \Im (f_{2l}) = 0.
	\label{eq:Im(f_l)}
\end{equation}

\section{}
\label{app:BdG}

In this appendix we will derive some properties of the bound states localized around a small ($L \ll \xi_S$) spherically symmetric impurity, suppressing the pairing potential. First, we prove that such impurity supports bound states with arbitrary large orbital momenta $l$. The energies $E_l$ of such states may be determined from the Bogoliubov-de Gennes equations\cite{Kopnin-book}
\begin{equation}
	({\cal{\hat{H}}}_l - E_l \check{1}) \left(
	\begin{array}{c}
	  u(r) \\
		v(r)
	\end{array} \right) =0,
	\label{eq:BdG}
\end{equation}
where
\begin{equation}
  {\cal{\hat{H}}}_l = \left(
	\begin{array}{cc}
	  H_l(r) & \Delta(r) \\
		\Delta(r) & -H_l(r)
	\end{array} \right),
  \label{eq:Hl_matrix}
\end{equation}
\begin{equation}
	H_l(r) = -\frac{\hbar^2}{2m} \frac{\partial^2}{\partial r^2} + \frac{\hbar^2 l (l+1)}{2mr^2} + U(r).
	\label{eq:Hl}
\end{equation}
The boundary conditions at the origin are $u(0) = v(0) =0$. The energy of the ground state may be estimated using the variational principle
\begin{equation}
	E_l^2 \leq \frac{\int_0^{\infty}{(u_T^*(r) \  v_T^*(r)) {\cal{\hat{H}}}_l^2 \left(
	\begin{array}{c}
	  u_T(r) \\
		v_T(r)
	\end{array} \right)dr}}{\int_0^{\infty} \left( \abs{u_T(r)}^2 + \abs{v_T(r)}^2 \right) dr},
	\label{eq:variational}
\end{equation}
where $u_T(r)$ and $v_T(r)$ are some trial functions satisfying the boundary conditions. Let us take $u_T(r) = v_T(r) = u_l(r)e^{-\delta r}$, where $u_l(r)$ is defined in Eq. \eqref{eq:u_l}, and $\delta$ is an adjustable parameter. According to Eq. \eqref{eq:variational}, we have
\begin{eqnarray}
	& E_l^2 \leq \Delta_0^2 & \nonumber  \\
	& + \frac{\int_0^{\infty} \left[ \left( \frac{\hbar^2 \delta}{2m} \right)^2 \abs{\frac{du_l}{dr} - \delta u_l}^2 + [\Delta^2(r) - \Delta_0^2] \abs{u_l(r)}^2 \right] e^{-2\delta r} dr}{\int_0^{\infty} \abs{u_l(r)}^2 e^{-2\delta r} dr}, &
	\label{eq:El_estim}
\end{eqnarray}
where $\Delta_0$ is the value of the gap far from the impurity. It can be seen that if there is a region with $\Delta(r)^2 < \Delta_0^2$ (and everywhere $\Delta(r)^2 \leq \Delta_0^2$), by taking a sufficiently small parameter $\delta$ one can make the second line of Eq. \eqref{eq:El_estim} negative. Thus, $E_l^2 < \Delta_0^2$, which means that a bound subgap state exists for arbitrary $l$.

Now we will derive a few explicit expressions for $E_l$. Of course, these energies can be determined by solving the Bogoliubov-de Gennes equations \eqref{eq:BdG}, but this is not necessary: at $L \ll \xi_S$ and $l^2 \ll k_F \xi_S$ this solution will yield Eq. \eqref{eq:El}, which will be analyzed in the following.

In the case when the electric potential $U$ is absent or negligible compared to the centifugal potential, the functions $u_l$ in Eq. \eqref{eq:El} can be expressed in terms of the spherical Bessel functions $j_l$\cite{Abramowitz}:
\begin{equation}
	u_l = 2k_F r j_l(k_F r)
	\label{eq:ul_centrifugal}
\end{equation}
(according to the definition \eqref{eq:u_l}, $u_l$ has a complex phase factor, but this does not matter here). 

We can compare our analytical result for $E_0$ with a numerical result obtained by Flatt\'e and Byers \cite{Flatte+97PRB}, who studied the eletctronic structure of defects of different nature. In particular, a local Gaussian suppression of the pairing potential on the scale of $k_F^{-1}$ has been considered. For an analytical estimate we assume that $\Delta(r)$ is proportional to the pairing potential, so that
\begin{equation}
	\Delta_1(r) = -\Delta_0 e^{-k_F^2 r^2}.
	\label{eq:Delta1_Gauss}
\end{equation}
Such approximation should work well when $L/\xi_S \ll 1$, so that perturbations of the Green functions at imaginary energies are small. For $k_F \xi_S = 10$ from Eqs.~\eqref{eq:f12l}, \eqref{eq:El} and \eqref{eq:ul_centrifugal} we obtain ${\cal E}_0 = 6\times 10^{-4} \Delta_0$. Our value for ${\cal E}_0$ appears to be one order of magnitude smaller than the numerical value given in Ref. \onlinecite{Flatte+97PRB}. The reason for this deviation is not very clear. The states with $l \neq 0$ have not been detected in the mentioned paper, so we may suppose that a higher numerical accuracy is required to calculate the bound state energies when they are lying very close to the gap edge.

Let us turn to the case of relatively large defects, when the energies $E_l$ can be calculated quasiclassically. Using Eq. \eqref{eq:Bessel_Bessel} and Debye's asymptotic expansion for the Bessel functions\cite{Jahnke_Funktionen}, we find that
\begin{equation}
	j_l(x) \approx \frac{1}{\sqrt{x z(x)}} \cos \chi(x),
	\label{eq:Debye}
\end{equation}
where
\begin{equation}
	z(x) = \sqrt{x^2 - \left( l + \frac{1}{2} \right)^2},
	\label{eq:z(x)}
\end{equation}
$\chi(x)$ is the quasiclassical phase, and it is required that
\begin{equation}
	z(x) \gg 1, \qquad z^3(x) \gg \left( l + \frac{1}{2} \right)^2.
	\label{eq:z_conditions}
\end{equation}
Note that $z(x) = 0$ approximately corresponds to the classical turning point, where the centrifugal potential equals the Fermi energy. When integrating in Eq.~\eqref{eq:f12l}, we replace $\cos^2 \chi(x)$ by $1/2$, which is its average over an oscillation period:
\begin{equation}
	f_{2l} \approx - \frac{k_F}{2\mu} \int_{k_F^{-1} \left( l+ \frac{1}{2} \right)}^L \frac{k_F r \Delta_1(r) dr}{\sqrt{k_F^2 r^2 - \left( l + \frac{1}{2} \right)^2}}.
	\label{eq:El_quasiclassical}
\end{equation}
It can be seen that the quasiclassical approach does not allow to determine the energies of the impurity states with large momenta ($l \gtrsim k_F L - 1/2$). Below we will demonstrate that these energies are exponentially close to the gap edge.

Previously, Gunsenheimer and Hahn\cite{Gunsenheimer+96PhysB} calculated the energies of the Andreev states on a normal sphere in a superconducting continuum (which corresponds to a step-like profile of $\Delta_1(r)$: $\Delta_1(r) = -\Delta_0$ when $r<L$, and $\Delta_1 = 0$ when $r>L$). Their approach is essentially quasiclassical, so their results should be equivalent to Eqs. \eqref{eq:El_quasiclassical} and \eqref{eq:El} when $L \ll \xi_S$. Substituting a rectangular profile of $\Delta_1$ into Eq. \eqref{eq:El_quasiclassical}, we obtain
\begin{equation}
	f_{2l} = \frac{\Delta_0}{2\mu} \sqrt{k_F^2 L^2 - \left( l + \frac{1}{2} \right)^2}.
	\label{eq:sphere_quasiclassical}
\end{equation}
This agrees well with the result from Ref. \onlinecite{Gunsenheimer+96PhysB}. 
Taking $k_F L \gg l+ 1/2$, we obtain the estimate \eqref{eq:flS_estimate}.

\begin{figure}[htb]
	\centering
		\includegraphics[width = \linewidth]{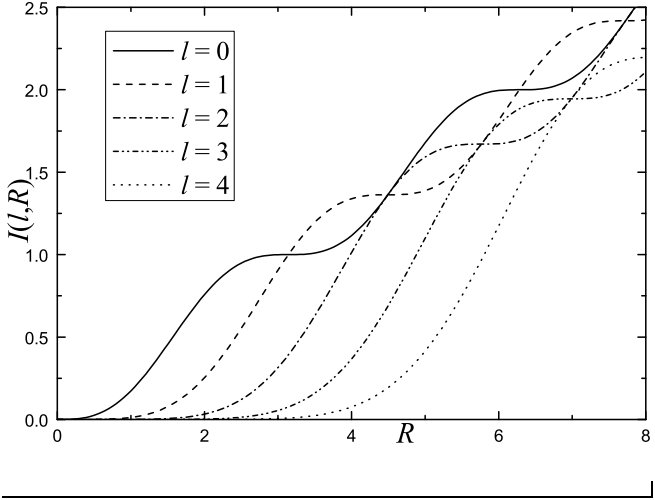}
	\caption{The functions $I(l,R)$ for several values of $l$.} \label{fig:Bessels}
\end{figure}

For smaller impurities or larger $l$ we need to go beyond the quasiclassical approximation. Using again a rectangular profile of $\Delta_1(r)$, we find that
\begin{equation}
	f_{2l} = \frac{\Delta_0}{\mu} I(l,k_F L),
	\label{eq:El_rectangular}
\end{equation}
where
\begin{eqnarray}
  & I(l,R) = \int_0^R x^2 j_l^2(x) dx = & \nonumber \\
	& \hspace{-0.7cm} = \! \frac{R^2}{2} \!\! \left\{ R[ j_l^2(R) + j_{l+1}^2(R)] \! - \! (2l+1) j_l(R) j_{l+1}(R) \right\} \!\!. &
	\label{eq:I}
\end{eqnarray}
The $I(l,R)$ vs. $R$ graphs for $l = 0..4$ are shown in Fig. \ref{fig:Bessels}. It can be seen that on the background of linear growth these functions exhibit oscillations with a period equal to $\pi$. These oscillations are a consequence of the discontinuity of $\Delta_1(r)$ at $r = L$. A smooth cross-over of $\Delta_1$ from $-\Delta_0$ to $0$ on a length scale larger than $k_F^{-1}$ will remove the oscillations.

In the limit of large $l$ -- $\sqrt{2(2l+3)} \gg k_F L$ -- one can replace the Bessel functions in Eq. \eqref{eq:I} by the first term of their Taylor series. This yields
\begin{equation}
	I(l,R) \approx \frac{R^{2l+3}}{(2l+1)!! (2l+3)!!}.
	\label{eq:I_bigl}
\end{equation}
Since
\begin{eqnarray}
  & (2l+1)!! = \frac{(2l+1)!}{2^l l!} \approx \exp{ \left[ -1 - l(1 + \ln 2) + \frac{1}{2} \ln \left( 2 + \frac{1}{l} \right) \right.} & \nonumber \\
	&  \left.+ (2l+1) \ln (2l+1) - l \ln{l} \right] &
	\label{eq:Stirling}
\end{eqnarray}
(see Stirling's formula in Ref. \onlinecite{Abramowitz}), the energies of the localized states approach the gap edge exponentially fast with growing $l$.

\section{}
\label{app:Band}

In this Appendix we calculate the width of the impurity band at a small concentration of pairing-potential impurities in the limit of strongly suppressed anisotropy, when the Green function satisfies Eq. \eqref{eq:universal1.5}. We note that in the vicinity of the energy of the bound state ${\cal E}$, given by Eq. \eqref{eq:E0_aniso}, the parameter $x = 1 - 2i f_{20} \mean{g_1}$ becomes small: $\abs{x} \ll 1$. Let us rewrite Eq. \eqref{eq:universal1.5} substituting $\mean{g_1} =-i (1-x)/(2f_{20})$:
\begin{widetext}
\begin{equation}
	\frac{E}{\mean{\Delta_0}} - 1 = -\frac{c_2 f_{20}}{\pi \nu_0 \mean{\Delta_0} x} - \frac{2f_{20}^2}{(1-x)^2} - \frac{\mean{\Delta'^2}}{f_{20} \hbar \mean{\Delta_0} \left(\tau_1^{-1} + \frac{\tau_2^{-1}}{x} \right)}(1-x).
	\label{eq:in_x}
\end{equation}
Now we will expand the right-hand side in the powers of $x$. We assume $\tau_2^{-1}/\abs{x} \ll \tau_1^{-1}$, which is valid at sufficiently small concentrations $c_2$, as we shall see further. Then, keeping terms up to the order of $x^2$, we obtain
\begin{equation}
	\delta E' + \left[ \frac{c_2 f_{20}}{\pi \nu_0 \mean{\Delta_0}} - \frac{\mean{\Delta'^2} \tau_1^2}{\hbar f_{20} \mean{\Delta_0} \tau_2} (1-x) \right] \frac{1}{x} + \left(4f_{20}^2 - \frac{\mean{\Delta'^2} \tau_1}{\hbar f_{20} \mean{\Delta_0}} \right) x + 6f_{20}^2 x^2 \approx 0,
	\label{eq:x_expanded}
\end{equation}
where $\delta E' = (E - \mean{\Delta_0} + {\cal E})/\mean{\Delta_0}$. Using the fact that $\tau_2^{-1} \leq 2c_2/(\hbar \pi \nu_0)$, it is easy to prove that 
\[ \frac{c_2 f_{20}}{\pi \nu_0 \mean{\Delta_0}} \gg \frac{\mean{\Delta'^2} \tau_1^2}{\hbar f_{20} \mean{\Delta_0} \tau_2} \abs{1-x}, \]
since the inequality \eqref{eq:bound_existence2} holds, and $\tau_1^{-1} \gg \sqrt[4]{\mean{\Delta'^2}}\sqrt{\mean{\Delta_0}}/\hbar$ in the limit of suppressed anisotropy. Also, it can be seen that the term proportional to $x^2$ can be neglected compared to the term proportional to $x$ when
\begin{equation}
	\abs{x} \ll 1 - \frac{\hbar}{8 \mean{\Delta_0} f_{20}^3 \tdep},
	\label{eq:x_applicability}
\end{equation}
where $\tdep^{-1} =2 \mean{\Delta'^2} \tau_1/\hbar^2$.
Then, we can determine $x$ from Eq.~\eqref{eq:x_expanded}:
\begin{equation}
	x = \frac{\frac{\delta E'}{4f_{20}^2} + \sqrt{ \left( \frac{\delta E'}{4f_{20}^2} \right)^2 - \frac{c_2}{\pi \nu_0 f_{20} \mean{\Delta_0}} \left( 1 - \frac{\hbar}{8 \mean{\Delta_0} f_{20}^3 \tdep} \right)}}{2 \left(\frac{\hbar}{8 \mean{\Delta_0} f_{20}^3 \tdep} - 1 \right)}.
	\label{eq:x}
\end{equation}
\end{widetext}
If the expression under the square root is negative at a given value $\delta E'$, then $x$ is complex, which means that the energy lies within the impurity band. This observation allows to determine the width $W$ of this band, which is given by Eq. \eqref{eq:band_uniform}.

\bibliographystyle{aipnum4-1}
\bibliography{Impurities}

\end{document}